\documentclass[preprint,superscriptaddress]{revtex4}
\usepackage{graphicx}
\usepackage{array}
\usepackage{amssymb}
\usepackage{amsfonts}
\usepackage{amsmath}
\usepackage{mathrsfs}
\usepackage{color}
\usepackage{booktabs}
\usepackage{threeparttable}
\usepackage{multirow}
\usepackage{subfigure}
\usepackage{times}
\usepackage{epsfig}
\usepackage{threeparttable}
\usepackage{chngpage}
\usepackage{latexsym}

\newcommand{\EqLabel}[1]{\label{#1}}
\newcommand{\EqRef}[1]{Eq. (\ref{#1})}
\def\eg{{e.g}. } 
\def\ie{{i.e}. } 
 
\def\etc{{etc}. }

\newcommand{\la}{\left\langle}
\newcommand{\ra}{\right\rangle}
\newcommand{\lbar}{\left|}
\newcommand{\rbar}{\right|}

\begin{document}



\newcounter{lastnote}
\newenvironment{scilastnote}{%
\setcounter{lastnote}{\value{enumiv}}%
\addtocounter{lastnote}{+1}%
\begin{list}%
{\arabic{lastnote}.}
{\setlength{\leftmargin}{.22in}}
{\setlength{\labelsep}{.5em}}}
{\end{list}}


\title{Interrelations among scientific fields and their relative influences revealed by an input-output analysis}


%

\author{Zhesi Shen}
\affiliation{School of Systems Science, Beijing Normal University, Beijing, 100875, P.R. China}

\author{Liying Yang}
\affiliation{National Science Library, Chinese Academy of Sciences, Beijing 100190, China}

\author{Jiansuo Pei}
\affiliation{School of International Trade and Economics, University of International Business and Economics, Beijing 100029, P.R. China}

\author{Menghui Li}
\affiliation{Beijing Institute of Science and Technology Intelligence, Beijing, 100044, P. R. China}

\author{Chensheng Wu}
\affiliation{Beijing Institute of Science and Technology Intelligence, Beijing, 100044, P. R. China}

\author{Jianzhang Bao}
\affiliation{School of Systems Science, Beijing Normal University, Beijing, 100875, P.R. China}

\author{Tian Wei}
\affiliation{School of Systems Science, Beijing Normal University, Beijing, 100875, P.R. China}

\author{Zengru Di}
\affiliation{School of Systems Science, Beijing Normal University, Beijing, 100875, P.R. China}

\author{Ronald Rousseau}
\affiliation{KU Leuven, Department of Mathematics, 3000 Leuven, Belgium}

\author{Jinshan Wu}\email{jinshanw@bnu.edu.cn}
\affiliation{School of Systems Science, Beijing Normal University, Beijing, 100875, P.R. China}


\date{}


\begin{abstract}
In this paper, we try to answer two questions about any given scientific discipline: First, how important is each subfield and second, how does a specific subfield influence other subfields? We modify the well-known open-system Leontief Input-Output Analysis in economics into a closed-system analysis focusing on eigenvalues and eigenvectors and the effects of removing one subfield. We apply this method to the subfields of physics. This analysis has yielded some promising results for identifying important subfields (for example the field of statistical physics has large influence while it is not among the largest subfields) and describing their influences on each other (for example the subfield of mechanical control of atoms is not among the largest subfields cited by quantum mechanics, but our analysis suggests that these fields are strongly connected). This method is potentially applicable to more general systems that have input-output relations among their elements.
\end{abstract}

\maketitle 


National science funding agencies and science policymakers often have to decide on which science or technology fields a nation will prioritize for a period of time. To answer this question, the funding agencies need to assess the (future) relative importance of all scientific fields. Furthermore, once the target, \ie the prioritized field, is chosen, the question of which other fields support the target field becomes an important consideration. 

These two questions are relevant not only to policymakers and committees in such agencies, but also to individual scientists, academic committees and university departments. Of course, one can apply peer review, relying on the opinions, feelings and visions of individual experts. With the rise of the era of big data, a natural question is whether technical analyses using large collections of published patents and research articles can help answer such questions.

The question of the relative importance of and influences between scientific fields has not yet been answered completely, admitting that investigating connections between scientific fields and technological sectors is one of the areas of investigation in the field of scientometrics\cite{JST_Indicator,Narin:Linkage}. In \cite{JST_Indicator}, the Japan Science and Technology Agency (JST) was interested in knowing, for a given sector of patents, which scientific fields have been the primary sources of published information. The simple approach used in \cite{JST_Indicator} is to calculate how journal articles cited in a specific sector of patents are distributed across all scientific fields. In \cite{Narin:Linkage}, the authors were more focused on how the patterns of citation between patents and scientific publications changed due to national origin and over time. Such analyses based on directly counting the number of articles, patents and citations, are referred to as direct analyses. In this simple, direct statistical approach, an indirect contribution from scientific fields to sectors of patents is missing: If there is one sector of patents $T_{\alpha}$, which heavily relies on one scientific field $S_{i}$, which in turn makes use of concepts and techniques from another scientific field $S_{j}$, then it is clear that even if there are no direct citations from $T_{\alpha}$ to $S_{j}$, $S_{j}$ is a major contributor to $T_{\alpha}$. These connections are referred as indirect connections. They are the main topics of this investigation. 

\begin{figure}[htbp]
\centering
\includegraphics[scale=0.8]{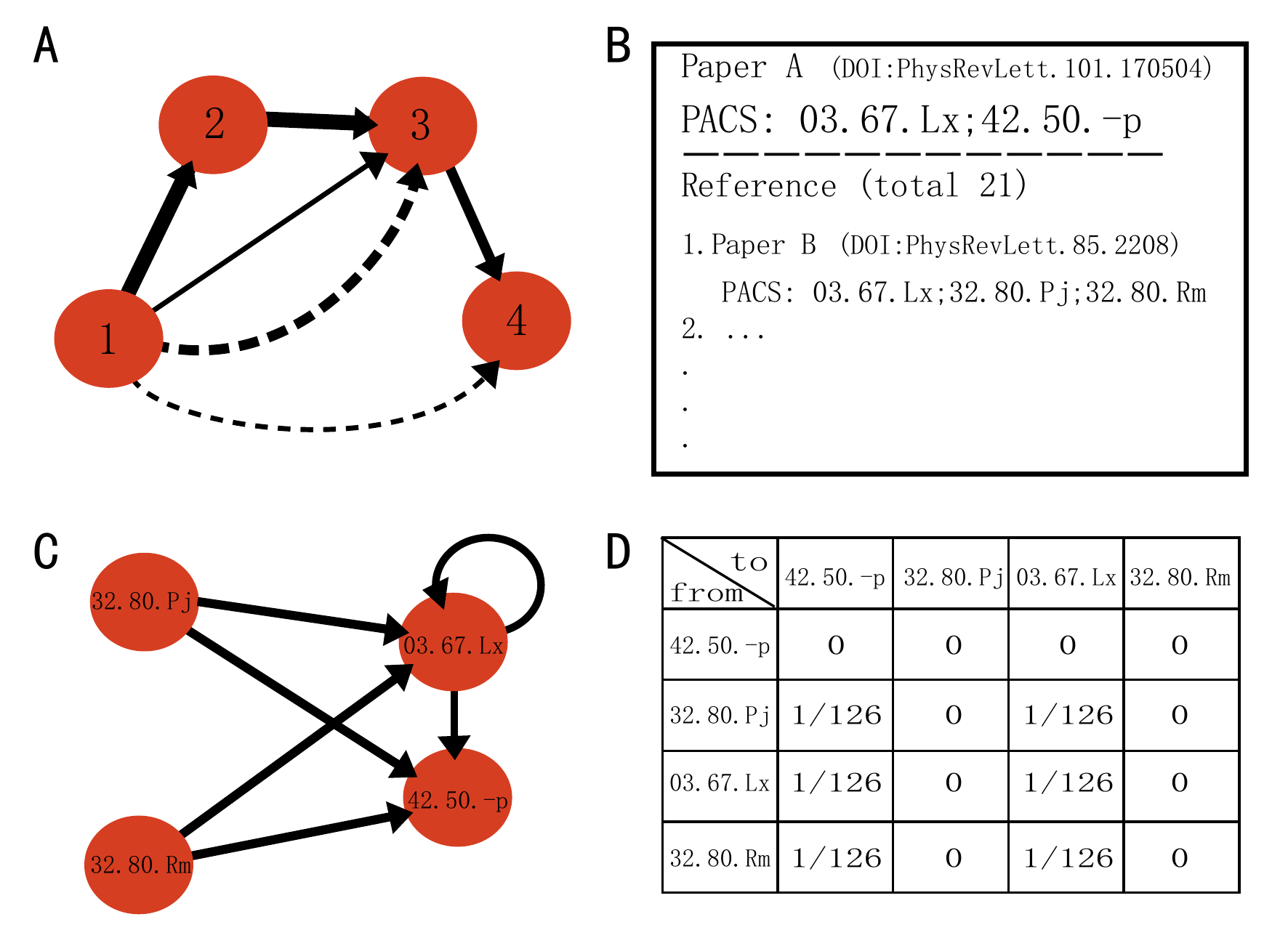}
\caption[Projecting citations among all APS papers into an input-output network/matrix of PACS codes]{Citations among all APS papers are converted into an input-output network/matrix of PACS codes. (A) A fictitious network in which there are relatively large differences between the direct and indirect influences between nodes. (B) A piece of the real APS citation network: paper A with PACS codes 03.67.Lx and 42.50.-p cites paper B with PACS codes 03.67.Lx, 32.80.Pj and 32.80.Rm. (C) In the corresponding input-output network of PACS codes, directed links from the PACS codes of Paper B to the PACS codes of paper A are added to the network of PACS codes following the citations from paper A to paper B. (D) A matrix version of B with numbers calculated using \EqRef{eq:weight}. }
\label{fig:illustration}
\end{figure}

This idea of considering direct as well as indirect relations, though straightforward, can not be underestimated. Results of such approaches are sometimes described as network effects\cite{WestVilhena:Network}. In Fig.\ref{fig:illustration}A, we illustrate an example of a citation relationship between scientific fields in which, indirect connections (between node $1$ and node $4$ or node $1$ and node $3$) could in principle play a more important role than direct ones, due to the lack of a direct connection between nodes $1$ and $4$ and a weak connection between nodes $1$ and $3$. While network science researchers, including those from social network analysis, have often used this perspective\cite{NetworkBook}, the network perspective is not yet a commonplace in scientometrics. This remark does not imply that scientometricians have not valued the network perspective \cite{WestVilhena:Network}. Indeed, the network effect is the key idea behind Google's PageRank algorithm\cite{BrinPage:PageRank} and its scientific predecessor, the Pinski-Narin influence methodology\cite{PinskiNarin:Influence, Franceschet:PRLIOA}. The PageRank algorithm has been used to measure the relative importance of journals\cite{West:EigenFactor} and articles\cite{Redner_PRonAPS,Ma_PR}.

Now that our work has been placed in its proper context, we first note that we will focus on scientific fields instead of journals and articles. Therefore, we may naively adopt the PageRank algorithm or equivalently the Pinski-Narin influence methodology for our study, by classifying publications into scientific fields. 

However, our interest goes beyond a measure of relative importance. We also want to know which fields support or are supported by a given field. Therefore, we consider the Leontief Input-Output Analysis (LIOA) in economics\cite{IOTextBook, Leontief:Book}. LIOA is a method of answering similar questions about economic sectors. In fact, the similarity between the ideas and motivations behind LIOA and PageRank has previously been described by Franceschet\cite{Franceschet:PRLIOA}. In LIOA, one starts with a direct input-output matrix $B$, where $b^{i}_{j}$ represents the number (or monetary value) of product $i$ required for producing one product $j$. Sector $N$, the last sector, is reserved for final consumers, so $b^{i}_{N}$ refers to the number (or value) of products from sector $i$ used per final consumer. This sector is also called final demands. Two typical questions in LIOA are as follows: First, what happens if the final demand increases? How will the total output of the other sectors change to match an increment in the demand for certain products; Second, which economic sector is the most important for the whole economy? What are the effects of removing one sector, \eg sector $i$, from the economy, on each of the other sectors in the economy? The former is usually discussed in terms of the Leontief inverse\cite{IOTextBook}, a solution to a specific linear equation while the latter is often discussed in terms of the so-called Hypothetical Extraction Method (HEM)\cite{Temurshoev:HEM}. Roughly speaking, in HEM people compare various quantities calculated in the complete LIOA and in the LIOA without sector $i$. In this way, if there is a large change in one of the quantities, \eg sector $j$'s output, sector $i$ is regarded as important for and especially influential on sector $j$.

Because these two questions concerning the relative importance of industrial sectors and their interrelations, such as the effect of changes in the output of product $i$ on product $j$, are very close to what we are interested in, we use the ideas of LIOA for the present study. To do so, we need to define an input-output matrix $B$ based on the citation relationships between scientific fields. Entries in $B$ could be, for example, the ratio between the number of citations from field $j$ to field $i$ and the total number of citations received by field $j$. In a sense, this ratio stands for the number of citations of papers in $i$ required for producing a citation in $j$. This provides a close parallel between LIOA and the problem we intend to study.

However, as we will show furtheron, this approach is not as straightforward as it may seem. New concepts and techniques are required to make LIOA applicable to study the scientometric problems that we are interested in. The key difference is that LIOA is performed on an open system, but the system of scientific fields is a closed system. There is not a natural external sector paralleling the final demand sector in economics unless, perhaps, if one includes patents. This would be a further step requiring more data than what we have at the moment. Thus, we need an input-output analysis method for closed systems. Furthermore, the number of citations is not a conserved quantity in the production of scientific works: the total number of citations received by a field is often not the same as the number of citations initiated from the field.

Fortunately, as we illustrate later, eigenvalues, which are the basis of our definition of Input-Output Factor (IOF), and eigenvectors, which are the basis of our definition of Input-Output Influence (IOI), are the key concepts we need for our closed-system input-output analysis. This relates our method to the PageRank algorithm or, equivalently, the Pinski-Narin influence methodology. 
Therefore, the method developed in this study - - an extension of LIOA for a closed system -- can also be regarded as an extension of the PageRank algorithm that makes it applicable to influences among the nodes in a network with an input-output relation. 

Aside from the methodological contributions toward answering the two questions we raised in the beginning, we find that, although overall our IOF is strongly correlated with the number of citations/publications, there are outliers in the correlation plots between the IOF and the number of citations/publications. Those outliers have either much stronger (\ie, Statistical Physics) or much weaker (\ie,  Relativity), influences on other fields when compared with the number of citations/publications in them. It seems to us that these outliers are intuitively understandable and plausible. Similar meaningful outliers have been identified in relational studies, in which  influences on and from individual fields are considered. For example, we found that 03 (QuanMech) is closely related to 37 (Mechanical control of atoms) while direct citations between the two are not significant. This demonstrates that our network-based analysis can go beyond studies based on direct statistics using the number of citations/publications.

We present the main idea and the formulae in the next section. After that, in $\S$\ref{sec:results}, we use a closed-system analysis to investigate relationships between the subfields of physics using records from the American Physical Society (APS) of published journals articles and discuss the validity of the information revealed by our analysis. A more general discussion of the validity of our closed-system input-output analysis can be found in $\S$\ref{sec:conclusion}. Discussions of some technical issues of our method and some additional results are reported in the Supplementary Materials.\\ 
 
\noindent
{\bf \large{Results}} \\
\noindent
{\bf Modified closed system input-output analysis(MCSIOA): the core idea.}
\label{sec:idea}
We will first summarize the open-system LIOA in economics and then modify it to make it applicable to closed systems. In fact, the first input-output model\cite{Leontief:Book} that Leontief proposed was a closed-system model and only later he and the vast majority of his followers turned to an open-system analysis. Let us assume the whole economy has $N$ sectors and each sector is a component such as Agriculture, Mining, Textiles \etc Starting from a matrix $x=\left(x^{i}_{j}\right)_{N\times N}$ representing the number or monetary value of all products of sector $i$ that are required for producing the products of sector $j$, one defines a matrix of direct input-output coefficients
\begin{equation}
b^{i}_{j} = \frac{x^{i}_{j}}{X^{j}},
\end{equation} 
where $X^{j} = \sum_{k}x^{j}_{k}$. With these elements $b^{i}_{j}$, we obtain
\begin{equation}
X^{i} = \sum_{j}b^{i}_{j}X^{j} \Longrightarrow X = B X,
\EqLabel{eq:closed}
\end{equation} 
meaning that $X$ is an eigenvector of matrix $B$ with eigenvalue $1$, the largest eigenvalue of matrix $B$. For simplicity, we call the eigenvector corresponding to the largest eigenvalue the largest eigenvector. 

If we separate the final demand sector, say sector $N$, from the other sectors of an economy, and denote it as $x^{i}_{N}=y^{N}$, we have
\begin{equation}
X^{i} = \sum_{j=1}^{N-1}b^{i}_{j}X^{j} + y^{N} \Longrightarrow X^{\left(-N\right)} = \left(I-B^{\left(-N\right)}\right)^{-1} Y,
\EqLabel{eq:open}
\end{equation} 
where $X^{\left(-N\right)}$ is what remains of vector $X$ after its $N$th element is removed and, similarly, $B^{\left(-N\right)}$ is the matrix $B$ after its $N$th row and $N$th column are removed. 
The inverse matrix is known as the Leontief inverse, and is denoted as $L= \left(1-B^{\left(-N\right)}\right)^{-1}$. $L$ is also called the full input-output coefficient matrix because it takes into account not only the direct coefficients but also the indirect ones. This can be observed even more clearly if we rewrite $L$ as follows:
\begin{equation}
\Delta X = L \Delta y = \sum_{n} \left(B^{\left(-N\right)}\right)^{n}\Delta y ,
\EqLabel{eq:Leontief}
\end{equation} 
assuming $\Delta y $ is known. 

In addition to the question of the system's response to a change in the final demand, LIOA can be applied to measuring the relative importance of sectors and the influences among them. 
This is called the Hypothetical Extraction Method (HEM)\cite{Temurshoev:HEM}. The basic idea is that for a given $\Delta y$ (without the previous $j$th element), one can define
\begin{equation}
\Delta^{\left(-j\right)} X = L^{\left(-j\right)} \Delta y = \left(1-B^{\left(-N-j\right)}\right)^{-1}\Delta y,
\EqLabel{eq:HEM}
\end{equation}
where $B^{\left(-N-j\right)}$ is what remains of matrix $B$ after both the $j$th and the $N$th ($j\neq N$) row and column are removed. One then compares $\Delta X$ with $\Delta^{\left(-j\right)} X$. If they are quite different (or, specifically, the $k$th element differs), then the $j$th sector is essential to the economy (to the $k$th sector). One may say that the importance of sector $j$ to the economy and to each other sector is concealed in the difference between $L$ and $L^{\left(-j\right)}$.

Due to the difference in the time scales of producing next-generation labor and manufacturing other products, it is plausible to separate the sector of final consumers from the other industrial sectors. However, in principle the sector of final consumers is an intrinsic `manufacturing' sector of the economy because it provides labor and accepts products. Let us now turn to the closed-system approach to input-output analysis, in which it is neither necessary nor possible to treat one sector as external to the system.  

Thus, the linear equation technique is clearly no longer applicable to our closed-system input-output analysis, but we may study the largest non-negative eigenvector of $B$ and $B^{\left(-j\right)}$ as long as those matrices have such an eigenvector. Ideally, we would also like to expect that such a largest non-negative eigenvector is unique for a given matrix $B$ or $B^{\left(-j\right)}$. However, in principle this is not necessarily true although this is almost always the case in the following empirical analysis. We introduce a robust analysis by adding a perturbative term to matrices $B$ and $B^{\left(-j\right)}$ to make the values all positive just as is used in the PageRank algorithm. Details are provided in the Supplementary Materials. For simplicity of notation, we still call those perturbed positive matrices $B$ and $B^{\left(-j\right)}$, of which each has a unique all positive largest eigenvector. 

We then consider the difference between the eigenvalues and eigenvectors of $B^{\left(-j\right)}$ and $B$. This relies on another interpretation of \EqRef{eq:closed}: the vector $X$ can be regarded as the specific combination of products that, when supplied to the economy, results in one hundred percent of the input becoming the output, \ie, the economy operates at full efficiency because the corresponding eigenvalue is $1$ and it is the maximum eigenvalue. Similarly, the maximum eigenvalue and the corresponding eigenvector of $B^{\left(-j\right)}$ are associated with the highest efficiency and the corresponding combination of products for the economy without sector $j$. Imagine the case in which sector $j$ has hardly any connections to other sectors, \ie, the values in the $j$th row and/or column are very small compared with other elements of $B$. Denoting the largest eigenvalue of matrix
$B^{\left(-j\right)}$ by $\lambda^{\left(-j\right)}$, then, $\lambda^{\left(-j\right)}$ will be very close to $1$. Otherwise, when elements in the $j$th row and column are relatively large, $\lambda^{\left(-j\right)}$ will be much smaller than $1$. The fact that all eigenvalues of the matrix $B^{\left(-N\right)}$ (and also all $B^{\left(-j\right)}$) must be less than or equal to $1$ in magnitude will be shown in the Supplementary Materials.

Therefore, we propose using the IOF defined by
\begin{equation}
S_{IO}^{j} = 1-\lambda^{\left(-j\right)}
\EqLabel{eq:IOfactorB}
\end{equation} 
to measure the relative importance of sector $j$. This answers the first question we raised in this paper. 

Let us now attempt to provide an answer to the second question. Intuitively, the influence of sector $j$ on each of the other sectors is concealed in the difference between $X$ and $\lbar \lambda^{\left(-j\right)}\ra$, which are respectively, the largest eigenvector of $B$ and $B^{\left(-j\right)}$. Thus, we propose the following quantity, which we call IOInfluence (IOI), to provide a comparison between $X$ and $\lbar \lambda^{\left(-j\right)}\ra$,
\begin{equation}
\Delta^{j}_{k} = \frac{\la  k \left. \rbar X \ra-\lambda^{\left(-j\right)}\left(\sum_{l\ne j} X^{l}\right)\la  k \left. \rbar \lambda^{\left(-j\right)} \ra}{\la  k \left. \rbar X \ra},
\EqLabel{eq:IOInfluenceB}
\end{equation}
where $\lbar\lambda^{\left(-j\right)}\ra$ is the largest eigenvector of matrix $B^{\left(-j\right)}$ and $\lbar k \ra$ is the column vector with all zeros except for the $k$th element. In a sense this eigenvector represents the best combination of products when sector $j$ is removed from the economy. The amount of total outputs of the new system without section $j$ intuitively should be $\lambda^{\left(-j\right)}$ times the original total output, thus the term $\lambda^{\left(-j\right)}\left(\sum_{l\ne j} X^{l}\right)$. Note that this definition of $\Delta^{j}_{k}$ is based on intuition and has not been fully justified.

\EqRef{eq:IOfactorB} and \EqRef{eq:IOInfluenceB} are the two core formulae in this paper. All of the calculations in the following sections are based on these two formulae. Within the general framework of the closed-system input--output analysis sketched above, we will now answer the two central questions raised at the beginning of this manuscript. \\

\noindent
{\bf MCSIOA applied to relationships between subfields in physics: the results}\\
\label{sec:results}
The above closed-system input-output analysis is now applied to relative importance of and influences among scientific fields. We consider subfields of physics as a case study.\\

\noindent
{\bf Construction of the closed Input-Output system.}
We use data regarding all papers published in APS (American Physical Society) journals between 1976 and 2013. A total of $390208$ papers have Physics and Astronomy Classification Scheme (PACS) codes. PACS is a classification system of subfields in physics consisting of $6$-digit $4$ to $5$-level codes. We will, however, use only the first $3$ levels. There are $10$(resp. $78$ and $937$) PACS codes at level $1$ (resp. level $2$ and level $3$). APS papers come with several author-defined PACS codes. The rich information encoded in such a classification system has been discussed in \eg \cite{Wu:HotTopics}. 

To establish the input-output system of subfields, we regard each PACS code as a sector. A citation received by a papers in one sector (PACS code $i$) from a paper in another sector (PACS code $j$) is modeled as an input from sector $i$ to sector $j$. We then count the papers and citations within the APS data. For example, if one paper $p$ published in sector $j$ cites a paper $q$ published in sector $i$, there is a link from $i$ to $j$. Each paper may have multiple PACS codes. For instance, if in a time window $t$, a paper $p$ having $P_{p}$ PACS codes, one of which is $j$, and cites $C_{p}$ papers, one of which is $q$, which has $P_{q}$ PACS codes one of which is $i$, then the contribution towards the input-output relation from $i$ to $j$ due to the citation from paper $p$ to paper $q$ is
\begin{equation}
x^{i}_{j}\left(p\rightarrow q\right) = \frac{1}{P_{p}P_{q}C_{p}}.
\EqLabel{eq:weight}
\end{equation} 
The time window we use in this study is five years. We provide an example of the weighted network in Fig.\ref{fig:illustration}, where a citation, as in Fig.\ref{fig:illustration}A, from Paper A to Paper B is converted into a network, as in Fig.\ref{fig:illustration}B, and a matrix representing the weighted network, as in Fig.\ref{fig:illustration}C, following \EqRef{eq:weight}. Input-output networks/matrices $\left(x^{i}_{j}\right)_{N \times N}$ of PACS codes can be established at various levels in this way. In LIOA in economics, $X^{i}=X_{i}$: the total input to an economic sector equals to the total output from that sector. 
Here it is not necessarily true that the citation count from the field is the same as the citation count to the field. Luckily for us, we do not need this to be the case for the analysis to work.\\

\noindent
{\bf The relative importance of subfields and its evolution.}
With the set of input-output networks/matrices ($\left(x^{i}_{j}\right)_{N\times N}$, and matrices $B$) of PACS codes for different time periods, we first discuss the relative importance of subfields and how this evolves.

\begin{figure}[htp]
\centering
\includegraphics[scale=0.4]{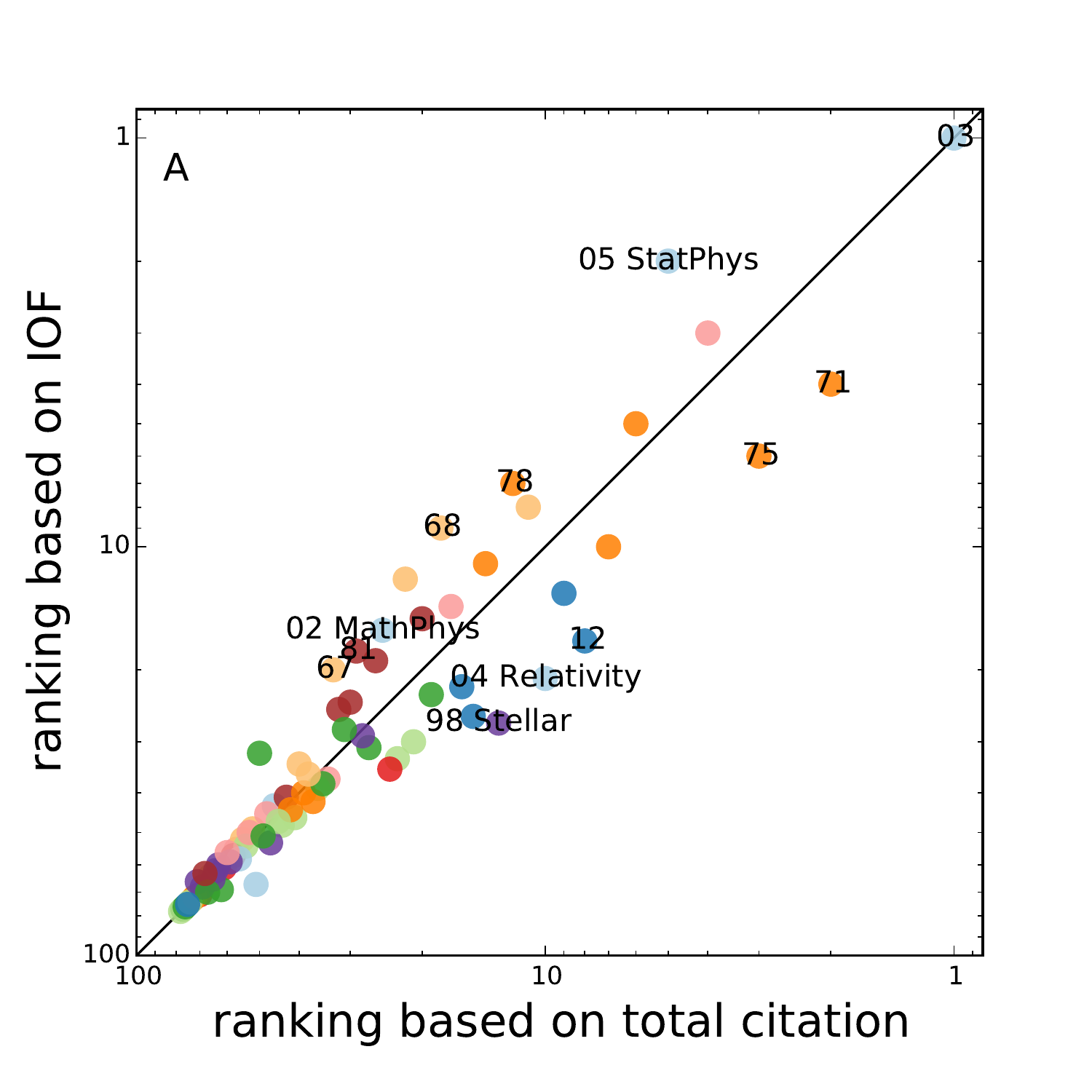}\includegraphics[scale=0.4]{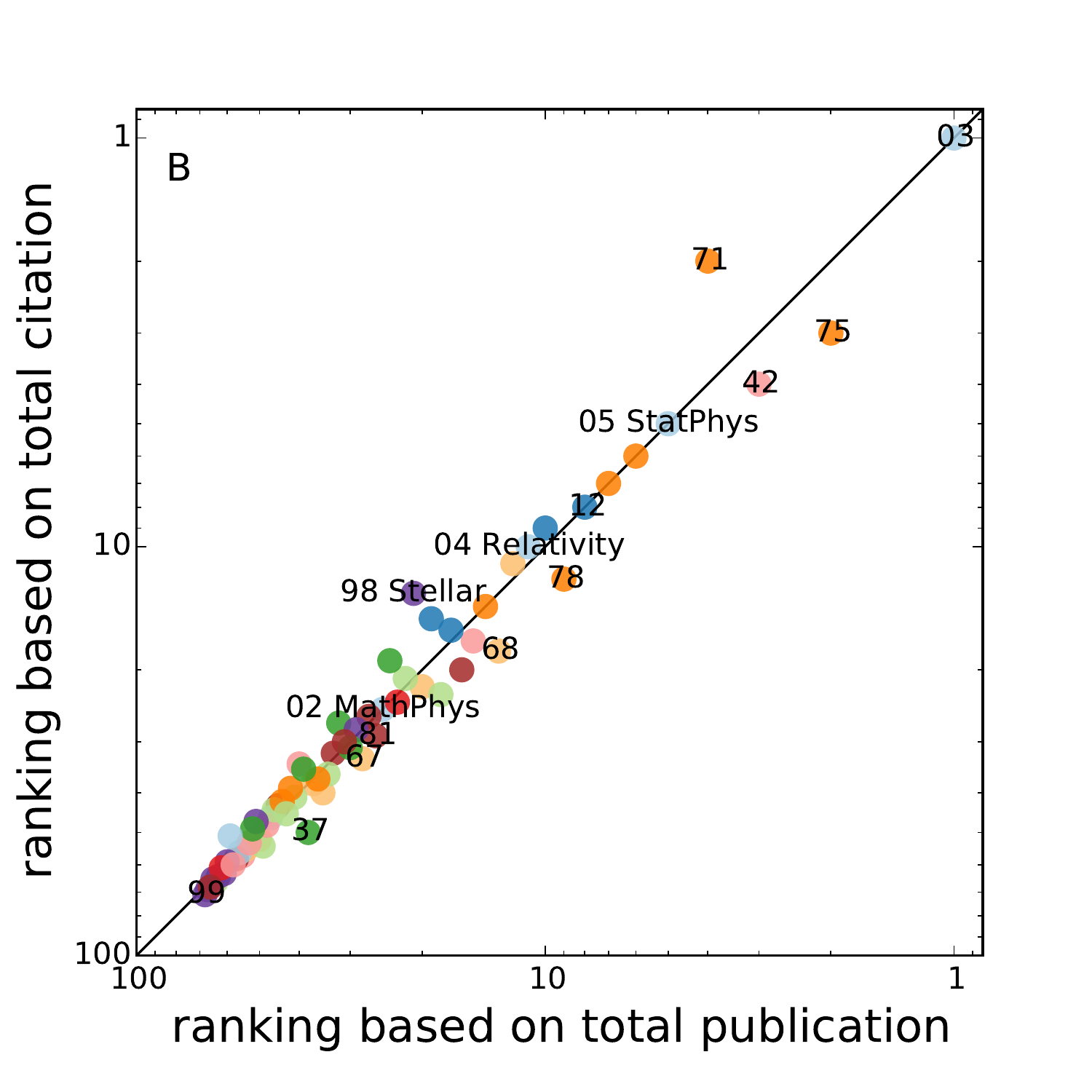}
\caption[Correlation between the citation rankings and IOF rankings of subfields]{(A)Correlation between the citation rankings and IOF rankings of subfields during the years between 2009 and 2013 is plotted in this figure. The $y$-axis ($x$-axis) represents rankings based on the IOF (total citations received) of each level-$2$ PACS code. In the region above the diagonal line along which the two rankings are equal, PACS codes have higher (smaller $y$ values, toward the top) IOF rankings than citation rankings, as is the case for 05 (StatPhys) and 02 (MathPhys). In the lower region, the citation rankings of the PACS codes are higher (smaller $x$ values, toward the right) than their IOF rankings, as is the case for 04 (Relativity) and 98 (Stellar). (B) To provide a comparison, we plot the number of citations received versus the number of publications. This considers only the direct connections between fields. The two numbers are highly correlated and those fields that stand out in (A) are no longer exceptional in this figure. See Supplementary Materials for these figures of sub fields at other levels.}
\label{fig:IOFvsCitation}
\end{figure}

First, we examine the correlation between the relative importance, as measured by the IOF, and by the number of times each subfield is cited. In Fig. \ref{fig:IOFvsCitation}A, we compare the IOF rankings of PACS codes with the rankings obtained from the total number of citations received by all papers with corresponding PACS codes. As shown in the figure, although the two rankings are  correlated, there are some outliers: some fields, such as 05 and 02, have relatively higher IOF rankings (smaller $y$ values, toward the top in the figure) whereas others, such as 04 and 98, have higher citation rankings (smaller $x$ values, toward the right in the figure). 

PACS 05 is the field of ``Statistical physics, thermodynamics and nonlinear dynamical systems'' (StatPhys for short). From the correlations for 2009 -- 2013 shown in Fig.~\ref{fig:IOFvsCitation}, we see that 05 has a large influence on other fields of physics relative to the number of citations it received, and this has been the case for this field for the past few decades (See Fig.~\ref{fig:IOFvsCitation_time} in the main text and Fig. S2 in the Supplementary Materials). This means that not only were papers in StatePhys (05) cited directly by many papers in other fields, but that 05 plays an important indirect role: Many other influential papers cited those papers who directly cited papers in 05 and so on. This picture of the importance of StatPhys is consistent with our own intuition that, in recent years, concepts, models and methods from statistical physics have been extensively used in other scientific fields. 

Similar but slightly different behavior can be observed for PACS 02, ``Mathematical methods in physics''. It has a relative low IOF ranking and total number of citations. However, considering its low number of citations, its IOF score is outstanding. This means that the total number of citations received directly by this field is not very high, but its indirect effect makes this field more important than the number of received citations suggests.

PACS 04 and 98 are among the fields that have higher citation rankings than their IOF rankings. This result does not imply that those fields are less important: it just means that they have smaller influence on other fields. It is understandable that each of these fields are more like a closed field of their own. Many physicists may not need to know much about stellar systems (98) to conduct their research.

We performed a similar comparison between the citation rankings and publication rankings of the subfields. We observed from Fig.~\ref{fig:IOFvsCitation}B that these rankings are better correlated than the previous pair of rankings, so that, generally speaking, the outliers in Fig.~\ref{fig:IOFvsCitation}B stand out less. Consider, for example, the subfields 04 and 05 in the two figures: they are quite different in Fig.~\ref{fig:IOFvsCitation}A while they are both on the diagonal line in Fig.~\ref{fig:IOFvsCitation}B. We want to emphasize that by including indirect connections, IOF rankings provides some more insightful and valuable information than citation rankings and the publication ranking (at least in this case) because the latter only consider direct connections.  

There are other outliers in the correlation figure, but we focused on some fields with which we have personal knowledge. The complete data set is provided in the Supplementary Materials for further examination. The results on parallel studies on level-$1$ and level-$3$ subfields are also reported in the Supplementary Materials. 

The same plot can be used to reveal the time evolution of the relative importances of the subfields. In Fig.~\ref{fig:IOFvsCitation_time}, we plot the values, instead of the rankings, of the IOF and citation counts of all subfields between 1996 and 2011. The trajectories of a few subfields (05, 03, 04, 32, 61, 68, 74, 78, 82, 98) are highlighted. The following two facts were interesting and surprising to us. First, for a very long time (before the year 2007) 05 (StatPhys) had a higher IOF than 03 (QuanMech), and second, that several subfields of 60 (Condensed matter I) and 70 (Condensed matter II) have decreasing IOFs even in cases of increasing citation counts. For example, the citation count of 74 (Superconductivity) increases while its IOF decreases. See Supplementary Materials for the figures of sub fields at other levels.\\

\begin{figure}[htp]
\centering
\includegraphics[scale=0.5]{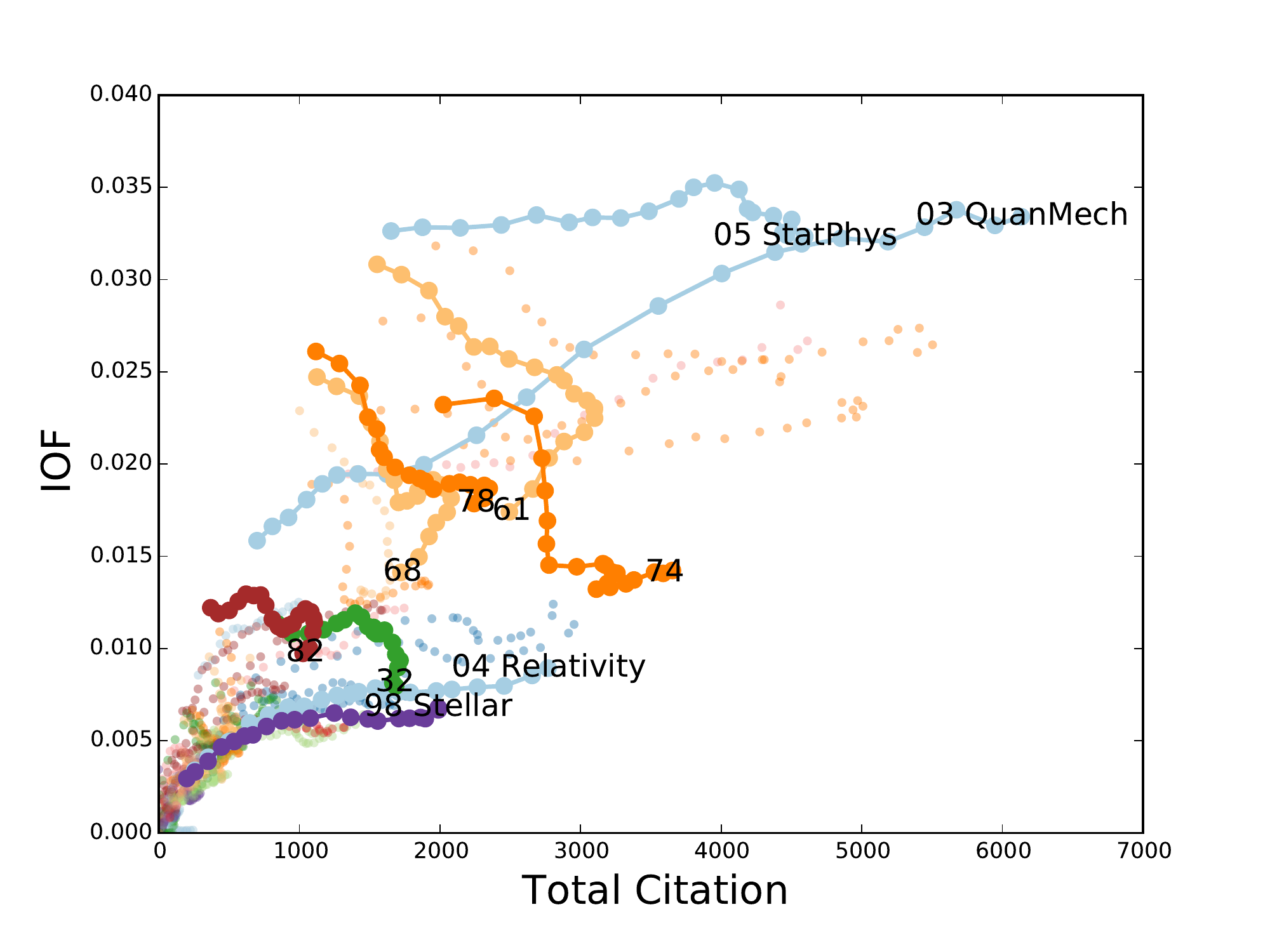}
\caption[Evolution of correlation between citations and the IOF values of subfields]{IOFs are plotted against citation counts for all subfields during each year from the year 1991 to 2011. Evolutions of sub fields 05, 03, 04, 32, 61, 68, 74, 78, 82, 98 are highlighted with their 2011 positions marked on this figure. Full data and a list of the top $20$ at all levels can be found in the Supplementary Materials.}
\label{fig:IOFvsCitation_time}
\end{figure}

\noindent
{\bf Influences among the subfields.}
For a given subfield $j$, we calculate $\Delta^{j}_{k}$. This describes how much the number of citations received by the subfield $k$ changes, directly and indirectly, if subfield $j$ is removed from the field of physics. Subfield $k$ relies strongly on subfield $j$ when $\Delta^{j}_{k}\ll 0$ and subfield $k$ can be regarded as a substitute for subfield $j$ when $\Delta^{j}_{k}\gg 0$.

\begin{figure}[htp]
\centering
\includegraphics[scale=0.8]{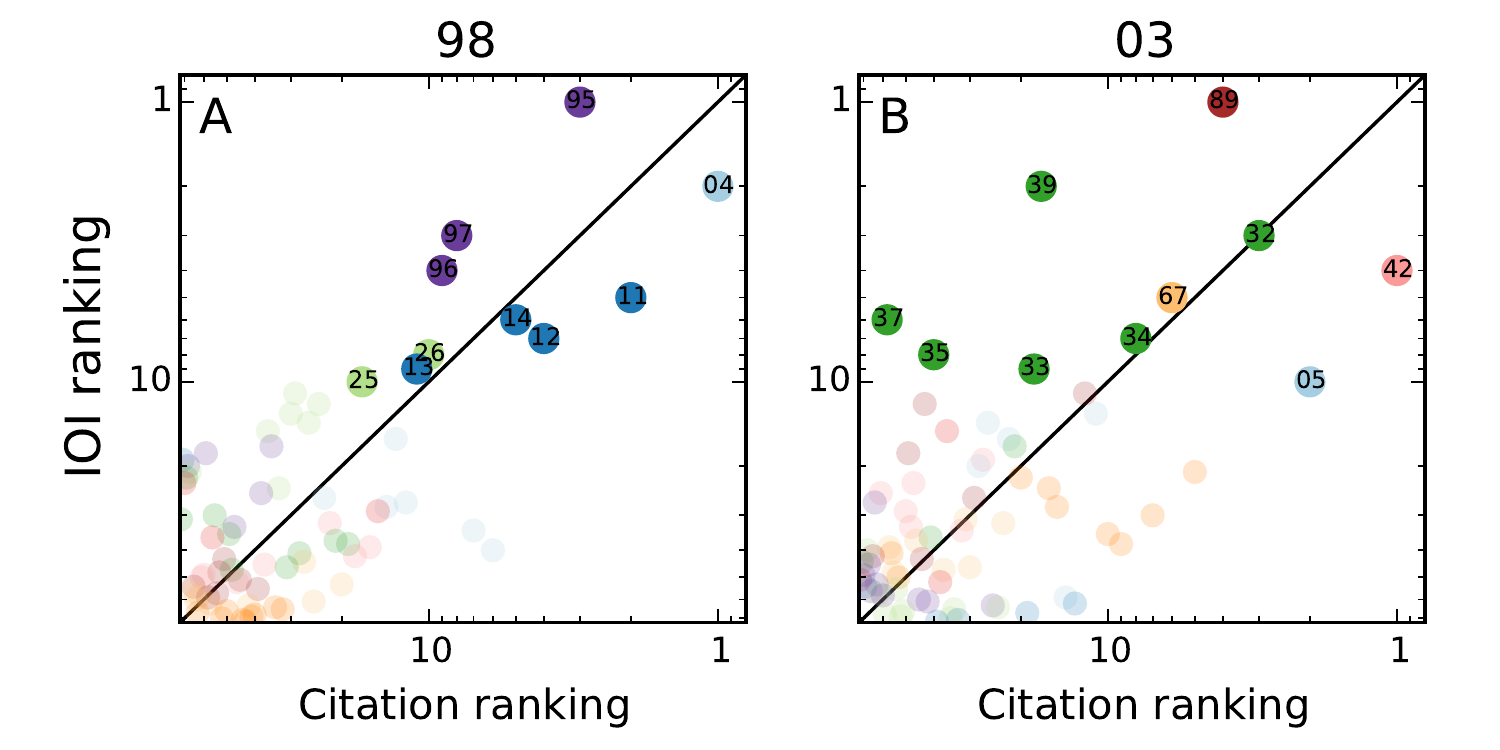}
\caption[Influences among the subfields]{Influences among the level-$2$ subfields. (A) For some fields (e.g., 98-Stellar systems), the influential rankings according to the IOI and citation counts are strongly correlated. It is also important to notice that for 98 the top $10$ closely related fields, most of which are in astronomy, relativity, stars etc., as indicated above, intuitively make sense. (B) For some fields (e.g., 03-QuanMech), there is a large difference between the two rankings. For example, while citation counts suggest that 37 (Mechanical control of atoms etc.) and 39 (
Instrumentation and techniques for atomic and molecular physics, later partially merged into 37) do not significantly depend on 03 (QuanMech), the IOI analysis suggests that 37 and 39 do in fact depend very much on 03. See Supplementary Materials for the figures of the remaining subfields at all levels.}
\label{fig:IOI}
\end{figure}

In Fig. \ref{fig:IOI} we use two specific subfields -- 98 (Stellar systems) and 03 (QuanMech) -- in the time interval 2004-2008 as examples. We see that there is a large difference between the influential sets, according to IOI and citation counts for subfield 03, while the difference is smaller for subfield 98. It is also important to note that, according to Fig.~\ref{fig:IOI}A, the top $10$ fields with the greatest influence on $98$ are generally in astronomy, relativity, stars, etc., which makes intuitive sense. This observation supports our intuitive definition of $\Delta^{j}_{k}$. From Fig. \ref{fig:IOI}B, we see that, if, for example, one wants to boost the development of 03, then it might be necessary to increase funding for 37 (Mechanical control of atoms etc.) and 39 (Instrumentation and techniques for atomic and molecular physics, later partially merged into 37), which are not in the top five fields cited from 03. A complete map of all the physics subfields at all levels is provided in the Supplementary Materials.\\

\noindent
{\bf \large{Conclusion and Discussion}}\\
\label{sec:conclusion}
In this paper we developed a method of closed-system input-output analysis and used it to study influences between subfields of physics using APS publication data. We found that by including both direct and indirect connections, our closed-system input-output analysis revealed deeper relationships among subfields than could be observed by directly looking at the numbers of citations and publications. This method provides an innovative approach to answering the two questions raised at the beginning of the paper: Given a set of fields, which is more influential thus should be supported preferentially? Given a specific priority, what other fields are necessary foundations for the targeted field and thus also need to be prioritized? When combined with time-series data, this method can also be used to track the development of the influences between scientific fields.   

Furthermore, the method proposed and developed in this work can be applied back to studies of economic systems and more generally to any type of networks with input-output relationships between the nodes. For example, a new type of influence factor of and among journals can be established based on this method. With more and more data available in this era of big data, it will be interesting to see more applications of this method. In addition, it will also be interesting to see a comparison between our results and the results from applying the PageRank algorithm to the same problem because both approaches consider indirect connections.





\noindent
{\bf \large{Acknowledgements}}\\
This work was supported by NSFC under Grant No. 61374175.\\

\noindent
{\bf \large{Author contributions}}\\
J.W., Z.D., J.P., L.Y., R.R. and C.W. designed this investigation, Z.S., 
J.B, T.W. and M.L. performed the analysis and J.W. Z.S. and R.R. also 
prepared the manuscript.

\noindent
{\bf Competing financial interests:} The authors declare no competing financial interests.\\

\noindent
{\bf Supplementary Materials}\\
Materials and Methods\\
Supplementary figures S1 to S9\\
Supplementary tables S1 and S2\\
Supplementary References (15-17)\\

\end{document}


\title{Supplementary Materials for Interrelations among scientific fields and their relative influence revealed by input-output analysis}

\maketitle
In these Supplementary Materials, we provide some extra explanation of our methods and some additional results which are mostly tables and figures, sometimes together with the data that are too large or too long to be included in the main text.

\section{Further details on methods and materials}
Uniqueness of the largest eigenvector of $B$ (and also $B^{\left(-j\right)}$) is a subtle and important technical issue of the analysis proposed in this work. Here we provide some further discussion on this issue.

\subsection{Uniqueness of the largest eigenvector of $B$ and $B^{\left(-j\right)}$} 

The Perron-Frobenius theorem of positive matrices, of which all elements are positive, states that each positive matrix has a unique eigenvector containing only positive values and the corresponding eigenvalue is the maximum real-value eigenvalue. Therefore, positive matrices have all of the good properties that we expect matrix $B$ and $B^{\left(-j\right)}$ to have. However, our matrices $B$ and $B^{\left(-j\right)}$ are not positive but only non-negative matrices. The Perron-Frobenius theorem of non-negative matrices, of which all elements are non-negative, claims that each irreducible non-negative matrix has a unique eigenvector containing only positive values and the corresponding eigenvalue is the maximum real-value eigenvalue. Note that matrix $B$ and $B^{\left(-j\right)}$ are not necessary irreducible. Due to this, the largest eigenvalue and the corresponding largest eigenvector might not be unique. Of course, it might be the case that the largest eigenvector is still all positive and it is unique. Thus, we performed the following additional analysis on matrix $B$ and all $B^{\left(-j\right)}$. 

First, we check the existence and uniqueness of this largest non-negative eigenvector in our practical calculations. After removing all sectors with no output ($X^{j}=0$) from matrix $X$ to define matrix $B$, we find that for all cases, such a largest non-negative eigenvector exists and it is unique for matrices $B$ and $B^{\left(-j\right)}$. However, although practically it is the case in our analysis, we can not guarantee that for other systems matrices $B$ and $B^{\left(-j\right)}$ always have this property.

Second, we check for irreducibility of matrices $B$ and $B^{\left(-j\right)}$ as that is required in the Perron-Frobenius theorem of non-negative matrices. One way to do that is to examine the strong connectivity of the graph corresponding to $B$ and $B^{\left(-j\right)}$. We have done so in this work using the non-recursive Tarjan's algorithm with Nuutila's modifications provided in the networkX software\cite{networkX} and find that at all PACS levels the strongly connected components of $B$ and $B^{\left(-j\right)}$ cover more than $96\%$ of all citations. At level $1$, for our $5$-year period analysis, the whole network $B$ is strongly connected already and all the corresponding networks of $B^{\left(-j\right)}$ are also strongly connected. At level $2$, the strongly connected subgraph of $B$ and $B^{\left(-j\right)}$, denoted as $\bar{B}$ and $\bar{B}^{\left(-j\right)}$, keeps $99\%$ of the citations in the whole network. At Level $3$, the citation network is relative sparse, so about $100$ sectors are excluded but the remaining strongly connected component keeps about $96\%$ of the citations. These large percentages means that even sometimes matrices $B$ and $B^{\left(-j\right)}$ might not be irreducible, they are very close to irreducible matrices. 

In principle, we can always identify and then focus on only the strongly connected $\bar{B}$ and $\bar{B}^{\left(-j\right)}$. This procedure is, however, quite demanding. Here we suggest to use a perturbative analysis. 

Third, we use a perturbative analysis to, in a sense, calculate the largest non-negative eigenvector directly from $B$ and $B^{\left(-j\right)}$ instead of from $\bar{B}$ and $\bar{B}^{\left(-j\right)}$. The following idea of this perturbative analysis comes from the PageRank algorithm and is quite straightforward: We want to compare the calculated largest eigenvectors of matrix $B$ and $B_{\alpha} = \left(1-\alpha\right) B + \alpha E$ with $\alpha$ as a numerical value being very close to $0$ and matrix $E$ is the matrix with every element being $1$. According to the Perron-Frobenius theorem of non-negative matrices, because $B$ might not be irreducible, the calculated largest eigenvector of $B$ might only be one of the a few of eigenvectors corresponding to the eigenvalues with the same maximum magnitude, while the calculated largest eigenvector of $B_{\alpha}$, since it is a positive matrix, is unique and corresponds to the largest eigenvalue, which is also unique. Now when we compare those two calculated eigenvectors, denoted as respectively $\lbar \lambda\left(B\right) \ra$ and $\lbar \lambda\left(B_{\alpha}\right) \ra$, it can be the case that the two vectors are rather different or that they are quite similar. Since $\lbar \lambda\left(B_{\alpha}\right) \ra$ is unique but $\lbar \lambda\left(B\right) \ra$ is not, in principle, the two vectors can be quite different even with $\alpha$ being very close to $0$: There are multiple $\lbar \lambda\left(B\right) \ra$s and they live in a multidimensional largest eigenvector space. Even with a tiny $\alpha$ the dimension of the the largest eigenvector space collapses into a one-dimensional one. This change of dimensions has a large effect unless the largest eigenvector space of $B$ is already one-dimensional. Therefore, we can find out whether the largest eigenvector space of $B$ is one-dimensional by simply looking at whether the following expression give a value numerically very close to $1$ or not,
\begin{align}
\mathcal{U} = \la \lambda\left(B\right) \rdot \lbar \lambda\left(B_{\alpha}\right) \ra (\alpha \approx 0).
\EqLabel{eq:innerU}
\end{align} 

We also want to compare eigenvectors of $\bar{B}$ and $B_{\alpha}$ since if we want to use the largest eigenvector of $B_{\alpha}$ then ideally we want this largest eigenvector to be close to the one from $\bar{B}$. Thus we define
\begin{align}
\mathcal{V} = \la \lambda\left(\bar{B}\right) \rdot \lbar \lambda\left(B_{\alpha}\right) \ra (\alpha \approx 0).
\EqLabel{eq:innerV}
\end{align} 

\begin{figure}[htp]
\centering
\includegraphics[scale=0.3]{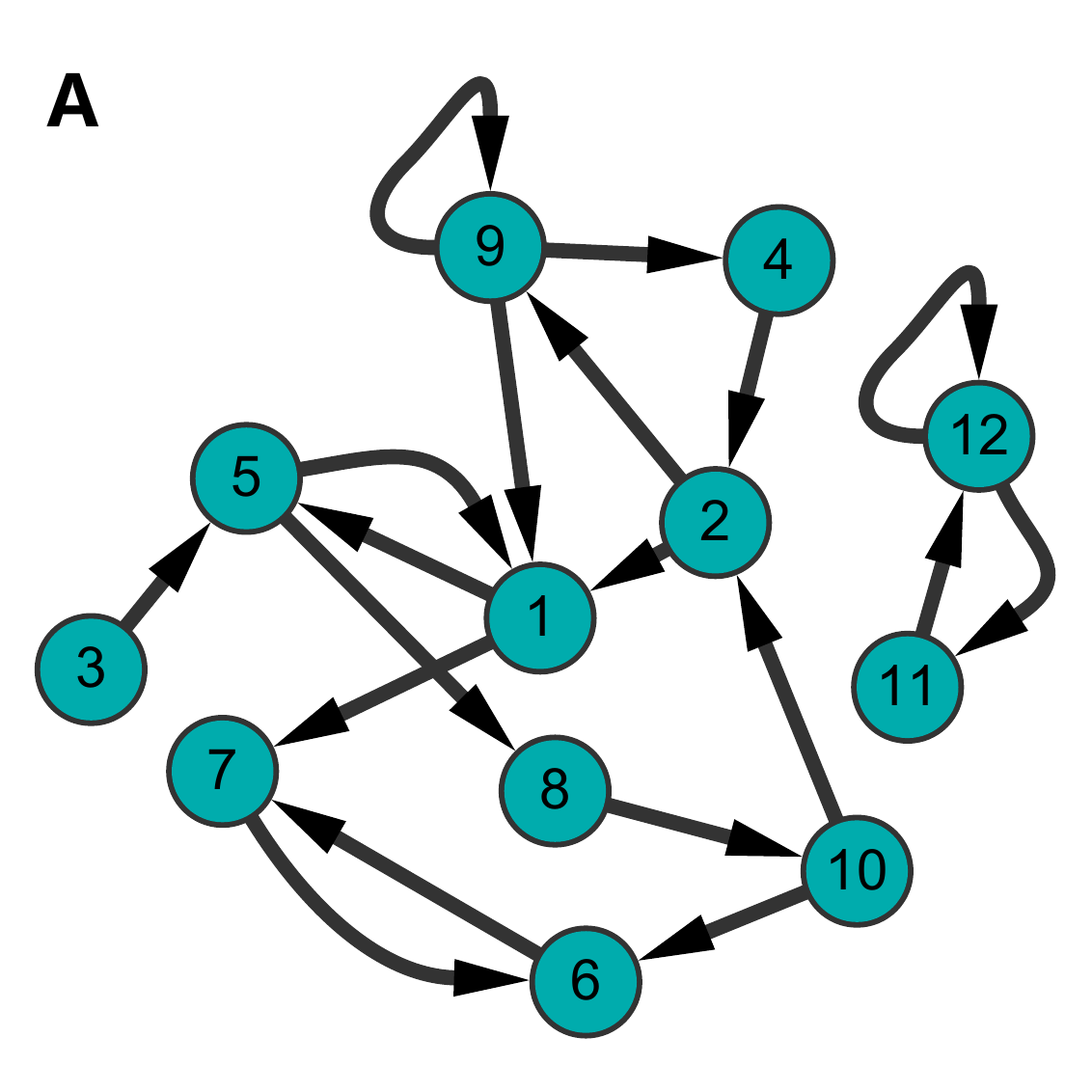}\includegraphics[scale=0.3]{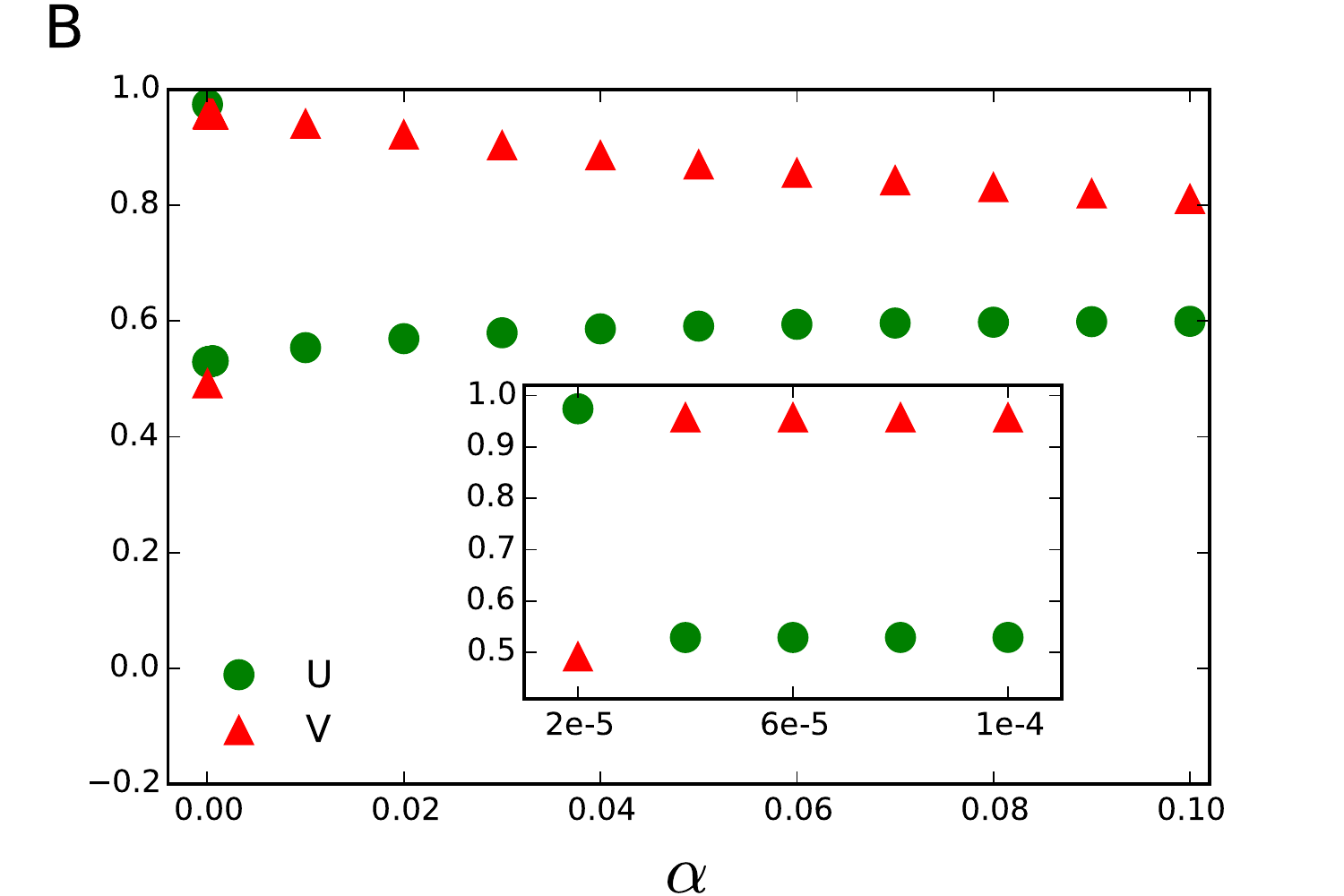}
\caption[Comparison between the calculated largest eigenvectors of $B$ and $B_{\alpha}$, and also $\bar{B}$ and $B_{\alpha}$ on a simple example]{(A) One example graph, which is not strongly connected. (B) $\mathcal{V}>\mathcal{U}$ for a wide regions of values of $\alpha$ except the case of $\alpha<0.00001$, where $\mathcal{U}>\mathcal{V}$.}
\label{fig:exampleUV}
\end{figure}

Note that ideally we expect that $\mathcal{V}$ to be close to $1$ and $\mathcal{U}$ to be slightly smaller than $1$. Theoretically, this holds for arbitrarily small $\alpha$ since introducing this $\alpha$ breaks the multiplicity of the largest eigenvalue in magnitude into a simple largest eigenvector. However, in numerical calculations, there is always a problem of finite accuracy so we use a simple example in Fig. \ref{fig:exampleUV} to estimate the sufficiently large value of $\alpha$. Our numerical calculation is performed with the Scipy~\cite{scipy} and specifically using the ARPACK linear algebra package~\cite{arpack} provided by the Scipy in Python. Remember that we do not want this value to be too large such that $\mathcal{V}$ becomes too small. 

The graph in Fig. \ref{fig:exampleUV} is not strongly connected and the corresponding adjacency matrix has multiple largest eigenvectors, \ie $\lbar \lambda\left(B\right) \ra$ is not unique. The calculated largest eigenvectors of $B$ ($B^{\left(-j\right)}$) and $B_{\alpha}$ ($B^{\left(-j\right)}_{\alpha}$) are compared and we find that for almost all values of $\alpha$, $\lbar\lambda\left(B_{\alpha}\right) \ra$ is closer to $\lbar\lambda\left(\bar{B}\right) \ra$ than $\lbar\lambda\left(B\right) \ra$ except when $\alpha<0.00001$. This means that even when the original matrix $B$ has multiple $\lbar\lambda\left(B\right) \ra$s introducing this extremely small $\alpha$ make $\lbar\lambda\left(B_{\alpha}\right) \ra$ to be unique and very close to the unique largest eigenvector of $\lbar\lambda\left(\bar{B}\right) \ra$. From what we have observed from this example, we use $B_{\alpha}$ and $B_{\alpha}^{\left(-j\right)}$ with $\alpha=0.00002$ instead of directly using matrix $B$ or $B^{\left(-j\right)}$ in all of our analysis presented in the main text. Again we want to emphasize that we intend to work on $\bar{B}$ and $\bar{B}^{\left(-j\right)}$, finding which is however very demanding, thus we instead turn to the much less demanding $B_{\alpha}$ and $B^{\left(-j\right)}_{\alpha}$. 

With such an extremely small $\alpha$, we regard this to be a simple technical issue without changing much the properties of the desired largest eigenvector of $\bar{B}$ and $\bar{B}^{\left(-j\right)}$. Due to this replacement, the core formulae Eq. (7) and Eq. (8) in fact need to be adjusted accordingly. However, since this $\alpha$ is extremely small and we do this for pure technical reasons, we regard the eigenvectors and eigenvalues to be those from matrix $B$ and $B^{\left(-j\right)}$ although they are not.

\subsection{Proof of $\lambda\left(B^{\left(-j\right)}\right)_{max} \leq 1$}

In this section, for simplicity, we assume that $B$ and $B^{\left(-j\right)}$ is irreducible. If $B^{\left(-j\right)}$ has one eigenvalue, whose magnitude is larger than $1$, then the corresponding eigenvector should also be the eigenvector of matrix $B$ (by adding simply $0$ at the $N$th component), thus matrix $B$ would have eigenvalue with magnitude larger than $1$. This conflicts with $1$ being the largest eigenvalue of $B$. Therefore, the magnitude of each eigenvalue of $B^{\left(-j\right)}$ must be less than or equal to $1$.

\section{Additional results}

\subsection{Tables and figures for influences of subfields of physics}
In Fig. 2 of the main text, we report correlations between ranks of level-$2$ subfields based on our IOF and number of citations. Here we provide the same correlation plots on level-$1$ and level-$3$ subfields of physics. 

\begin{figure}[htp]
\centering
\includegraphics[scale=0.23]{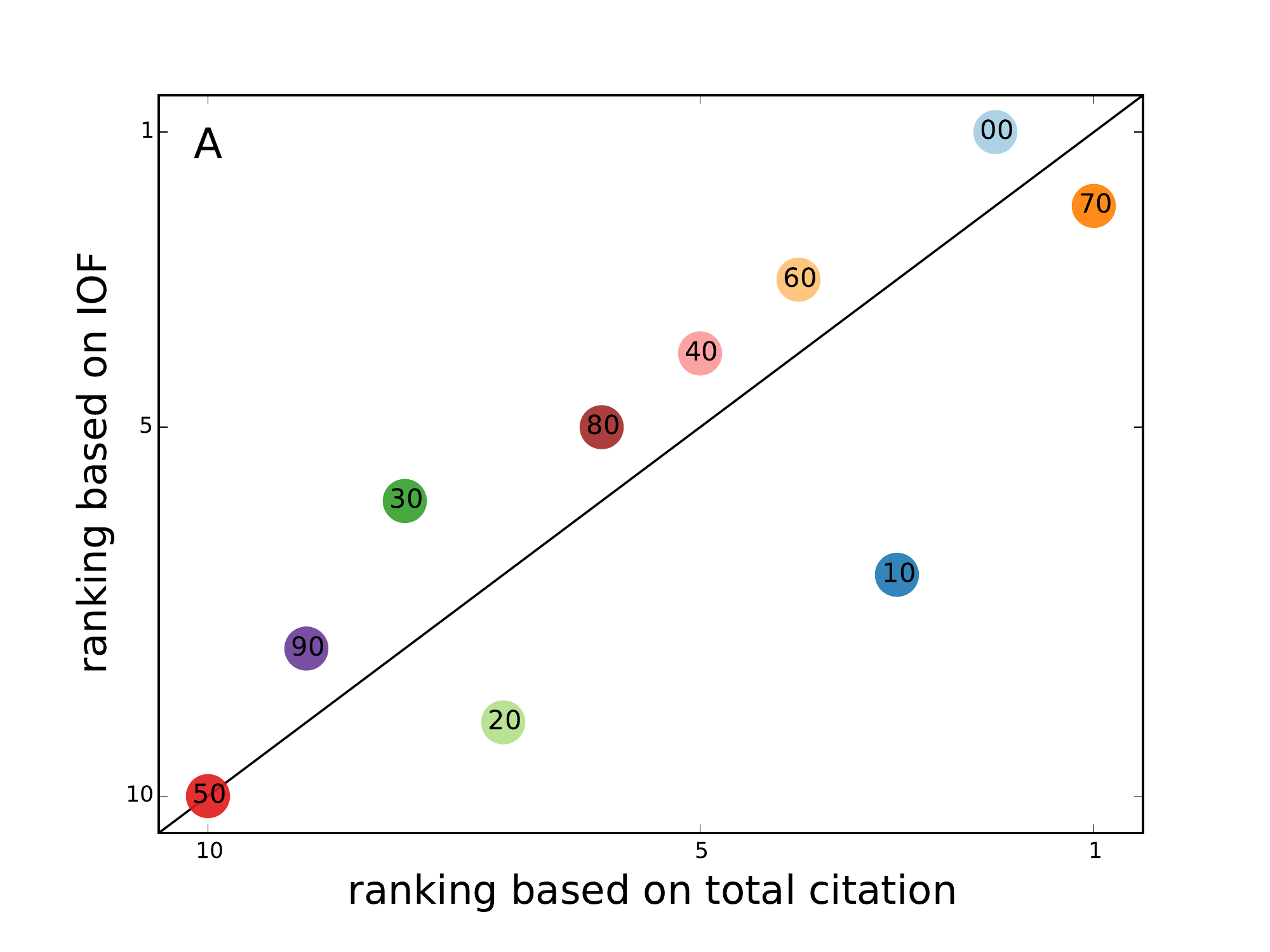}\includegraphics[scale=0.23]{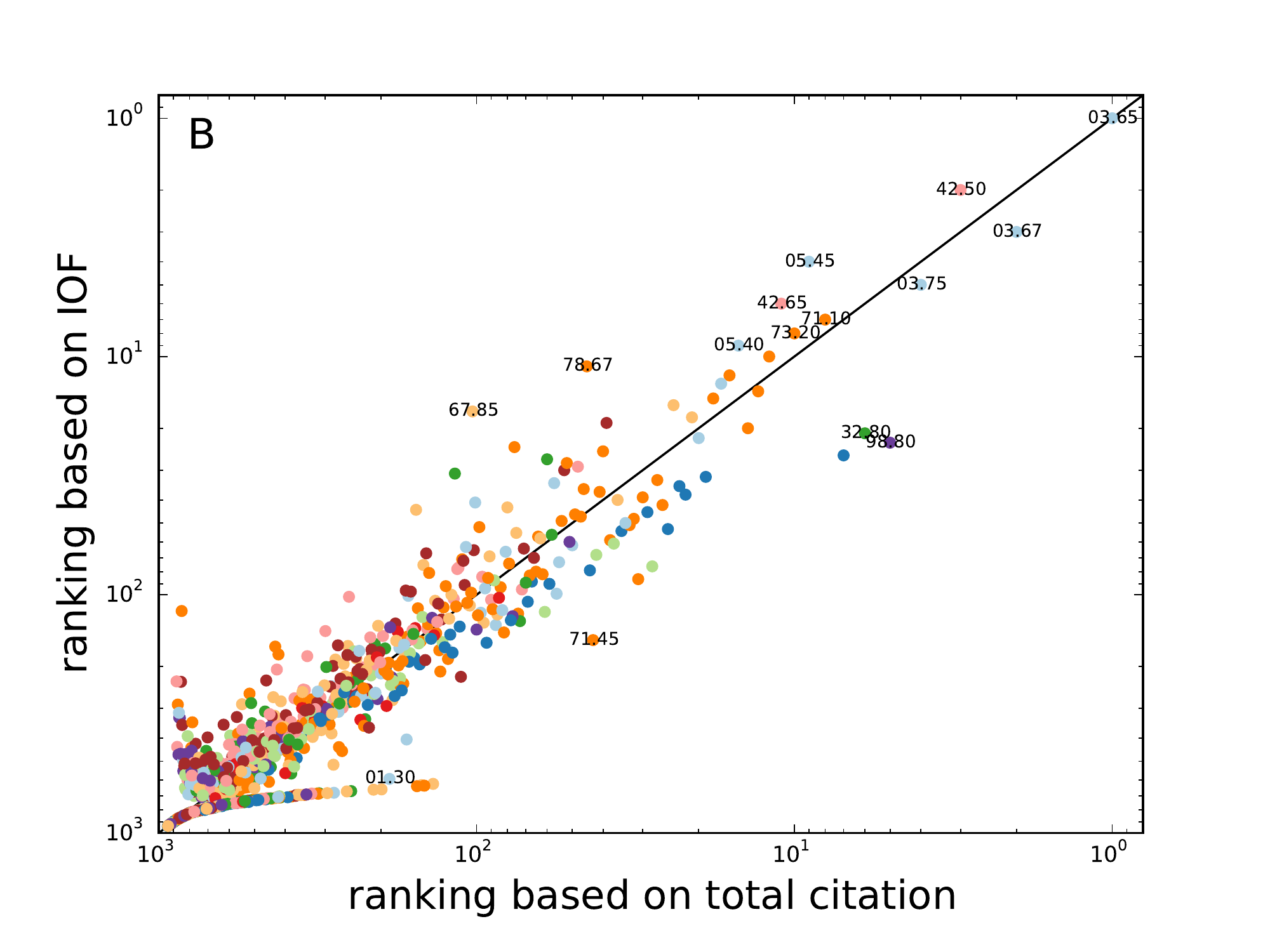}
\caption[Correlation between citations ranks and the IOF ranks of level-$1$ and level-$3$ subfields]{(A) Correlation between citations ranks and the IOF ranks of level-$1$ subfields is plotted in this figure. PACS codes such as 00(General Physics) have better IOF ranks than their citation ranks while citation ranks of PACS codes such as 10(The Physics of Elementary Particles and Fields) are better than their IOF ranks. (B) Correlation between citations ranks and the IOF ranks of level-$3$ subfields is plotted in this figure. Fields 67.85(Ultracold gases, trapped gases) and 78.67(Optical properties of low-dimensional, mesoscopic, and nanoscale materials and structures) have better IOF ranks than their citation ranks while fields 71.45(Collective effects) and 98.80(Cosmology) have better citation ranks than their IOF ranks.}
\label{fig:IOFvsCitation}
\end{figure}

We observed that there are again some outliers in the level-$1$ and level-$3$ correlation plots. At level-$1$, fields 00(General Physics) has better IOF ranks than their citation ranks while fields 10(The Physics of Elementary Particles and Fields) has better citation ranks than their IOF ranks. At level-$3$, fields 67.85(Ultracold gases, trapped gases) and 78.67(Optical properties of low-dimensional, mesoscopic, and nanoscale materials and structures) have better IOF ranks while fields 71.45(Collective effects) and 98.80(Cosmology) have better citation ranks. Field 30 at level-$1$ is rather special: Overall it has low IOF while it still have better IOF rank than the citation rank. This means that the influence of $30$ can be underestimated if judging from only number of papers and citations in the field although it is indeed not that influential.

To see better the time evolution of influences of subfields, we present an animated version of Fig. 3 in the main text.
\begin{figure}[htp]
\centering
\caption[Evolution of influences of level-$2$ subfields]{Evolution of influences of level-$2$ subfields between 1991 and 2011. Certain PDF reader such as the Acrobat PDF Reader might be needed to show to animation.}
\label{fig:annimate}
\end{figure}

We also include a table of the top $20$ most influential level-$2$ subfields at the year of 2011.
\begin{table}[htbp]\tiny
\caption{Top $20$ most influential level-$2$ subfields at the year of 2011. Full data can be downloaded at \revision{\url{subfield_list_level2.txt}}.}
\label{tab:level2}
\begin{center}
\begin{tabular}{ p{0.5cm} p{1.0cm} p{3cm} p{0.90cm}}
\hline
rank & pacs code & subfield & IOF \\
\hline
1  &  03 &  Quantum mechanics, field theories, and special relativity &  0.0334107413899  \\
2  &  05 & Statistical physics, thermodynamics, and nonlinear dynamical systems &  0.032334689129  \\
3  &  42 & Optics &  0.0266783205433  \\
4  &  71 & Electronic structure of bulk materials &  0.026051707365  \\
5  &  73 & Electronic structure and electrical properties of surfaces, interfaces, thin films, and low-dimensional structures &  0.0244555083998  \\
6  &  75 & Magnetic properties and materials &  0.0233384148198  \\
7  &  78 & Optical properties, condensed-matter spectroscopy and other interactions of radiation and particles with condensed matter &  0.0178594554773  \\
8  &  61 & Structure of solids and liquids; crystallography &  0.017412083493  \\
9  &  68 &  Surfaces and interfaces; thin films and nanosystems (structure and nonelectronic properties) &  0.0141285788655  \\
10  &  74 & Superconductivity &  0.0140740641051  \\
11  &  72 & Electronic transport in condensed matter &  0.0134113056464  \\
12  &  64 & Equations of state, phase equilibria, and phase transitions &  0.0131746830176  \\
13  &  11 &  General theory of fields and particles &  0.0124063117582  \\
14  &  47 & Fluid dynamics &  0.0121936938139  \\
15  &  87 & Biological and medical physics &  0.0120704923922  \\
16  &  02 & Mathematical methods in physics &  0.0118169903456  \\
17  &  12 & Specific theories and interaction models; particle systematics &  0.011308323537  \\
18  &  81 & Materials science &  0.0104836041511  \\
19  &  82 & Physical chemistry and chemical physics &  0.00973865082348  \\
20  &  67 & Quantum fluids and solids &  0.00948140209385  \\
\hline
\end{tabular}
\end{center}
\end{table}

\begin{table}[htbp]\tiny
\caption{Top $20$ most influential level-$3$ subfields at the year of 2011. Full data can be downloaded at \revision{\url{subfield_list_level3.txt}}. }
\label{tab:level3}
\begin{center}
\begin{tabular}{ p{0.5cm} p{1.0cm} p{3cm} p{0.90cm}}
\hline
rank & pacs code & subfield & IOF \\
\hline
1  &  03.65 & Quantum mechanics &  0.0148953211316  \\
2  &  42.50 & Quantum optics &  0.0147084409465  \\
3  &  03.67 & Quantum information &  0.0131002568588  \\
4  &  05.45 & Nonlinear dynamics and chaos &  0.0105192533981  \\
5  &  03.75 & Matter waves &  0.00982284895269  \\
6  &  42.65 & Nonlinear optics &  0.00822584929776  \\
7  &  71.10 & Theories and models of many-electron systems &  0.00779028463258  \\
8  &  73.20 & Electron states at surfaces and interfaces &  0.00771114964267  \\
9  &  05.40 & Fluctuation phenomena, random processes, noise, and Brownian motion &  0.00769361189889  \\
10  &  75.10 & General theory and models of magnetic ordering &  0.00728881536532  \\
11  &  78.67 & Optical properties of low-dimensional, mesoscopic, and nanoscale materials and structures &  0.00695251123923  \\
12  &  75.30 & Intrinsic properties of magnetically ordered materials &  0.00676542990058  \\
13  &  05.30 & Quantum statistical mechanics &  0.00670483397365  \\
14  &  71.15 & Methods of electronic structure calculations &  0.00668113108307  \\
15  &  75.50 & Studies of specific magnetic materials &  0.00617769198  \\
16  &  64.70 & Specific phase transitions &  0.00612260589561  \\
17  &  67.85 & Ultracold gases, trapped gases &  0.00611556368295  \\
18  &  64.60 & General studies of phase transitions &  0.00609626427723  \\
19  &  89.75 & Complex systems &  0.00594258830918  \\
20  &  74.25 & Properties of superconductors &  0.00592012932517  \\
\hline
\end{tabular}
\end{center}
\end{table}

We also include here evolutions of the top $20$ subfields at each level for the years 1991, 2001 and 2011. A text file of the full list of the subfields at each level for all the years between 1991 and 2011 is accessible through the following: Level-$1$ ($2$, $3$) list can be downloaded at \revision{\url{subfield_list_level1.txt} (\url{subfield_list_level2.txt}, \url{subfield_list_level3.txt})}.

\begin{figure}[htp]
\centering
\includegraphics[scale=0.4]{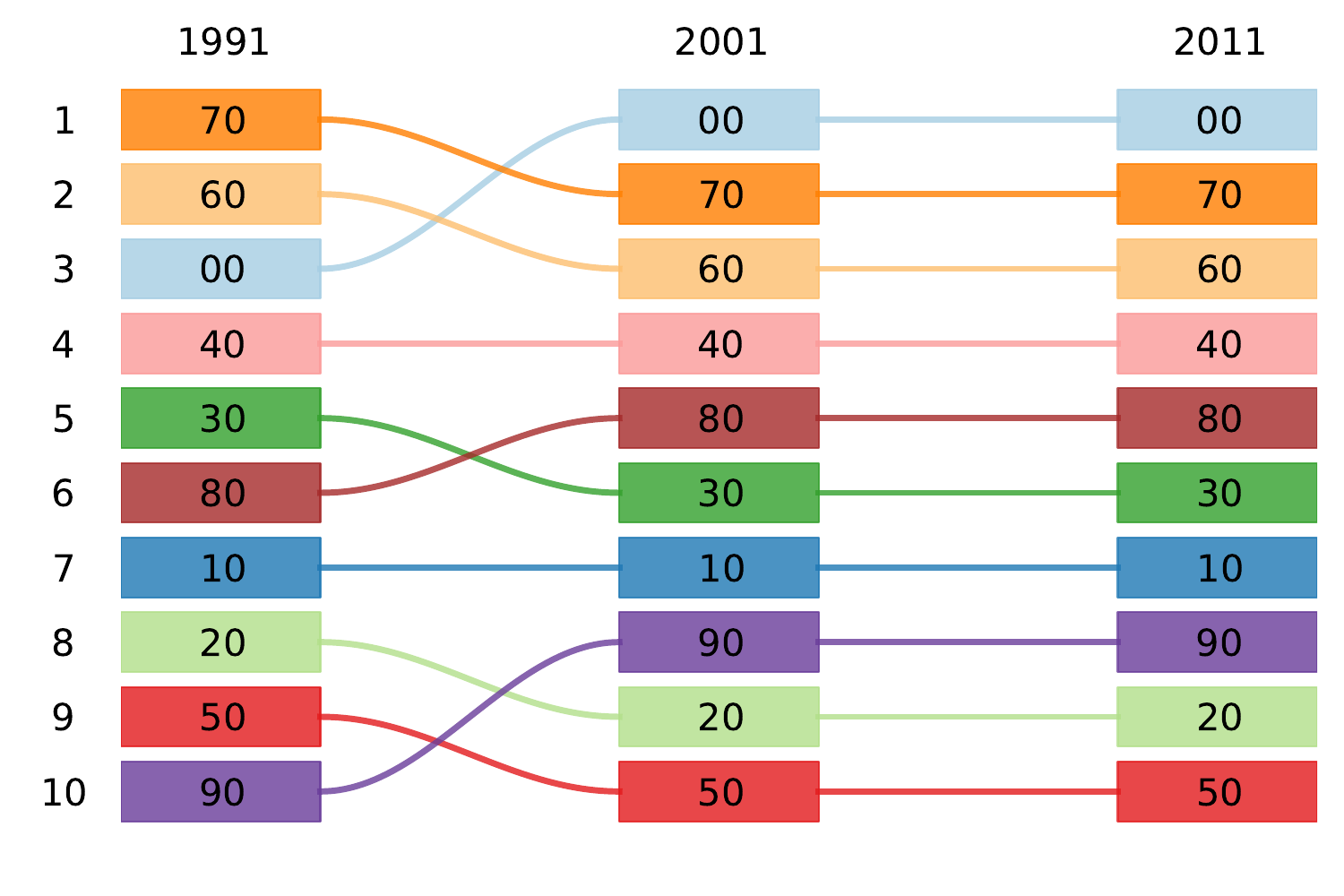}
\caption[Evolution of IOFs of all level-$1$ subfields]{Evolutions of all level-$1$ subfields for the years 1991, 2001 and 2011. Field 00 (General Physics) moved from the third in 1991 to the top in 2001 and stayed at the top till 2011. The ranks are quite stable after the year 2001. Full data can be downloaded at \revision{\url{subfield_list_level1.txt}}.}
\label{fig:evolution20-level1}
\end{figure}

\begin{figure}[htp]
\centering
\includegraphics[scale=0.4]{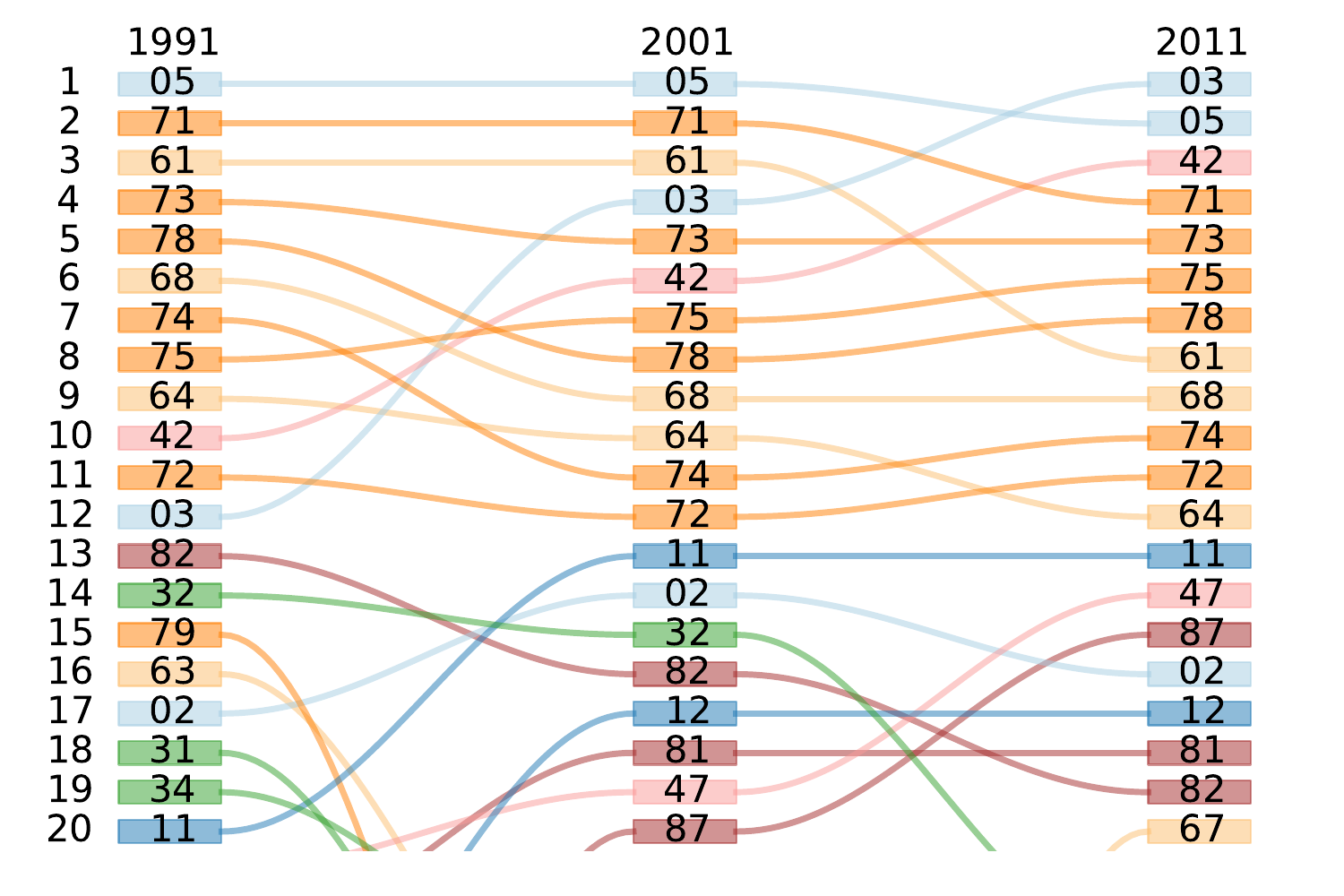}
\caption[Evolution of the top $20$ most influential level-$2$ subfields]{Evolutions of top $20$ most influential level-$2$ subfields for the years 1991, 2001 and 2011. Field 03 (QuanMech) has been increasing during the whole period and it has reached the top at 2011, in fact around 2009 as indicated in Fig. 3 in the main text. We also observed that overall influences of fields in 60 and 70 are stable or slightly decreasing compared to the year 1991. Full data can be downloaded at \revision{\url{subfield_list_level2.txt}}.}
\label{fig:evolution20-level2}
\end{figure}

\begin{figure}[htp]
\centering
\includegraphics[scale=0.4]{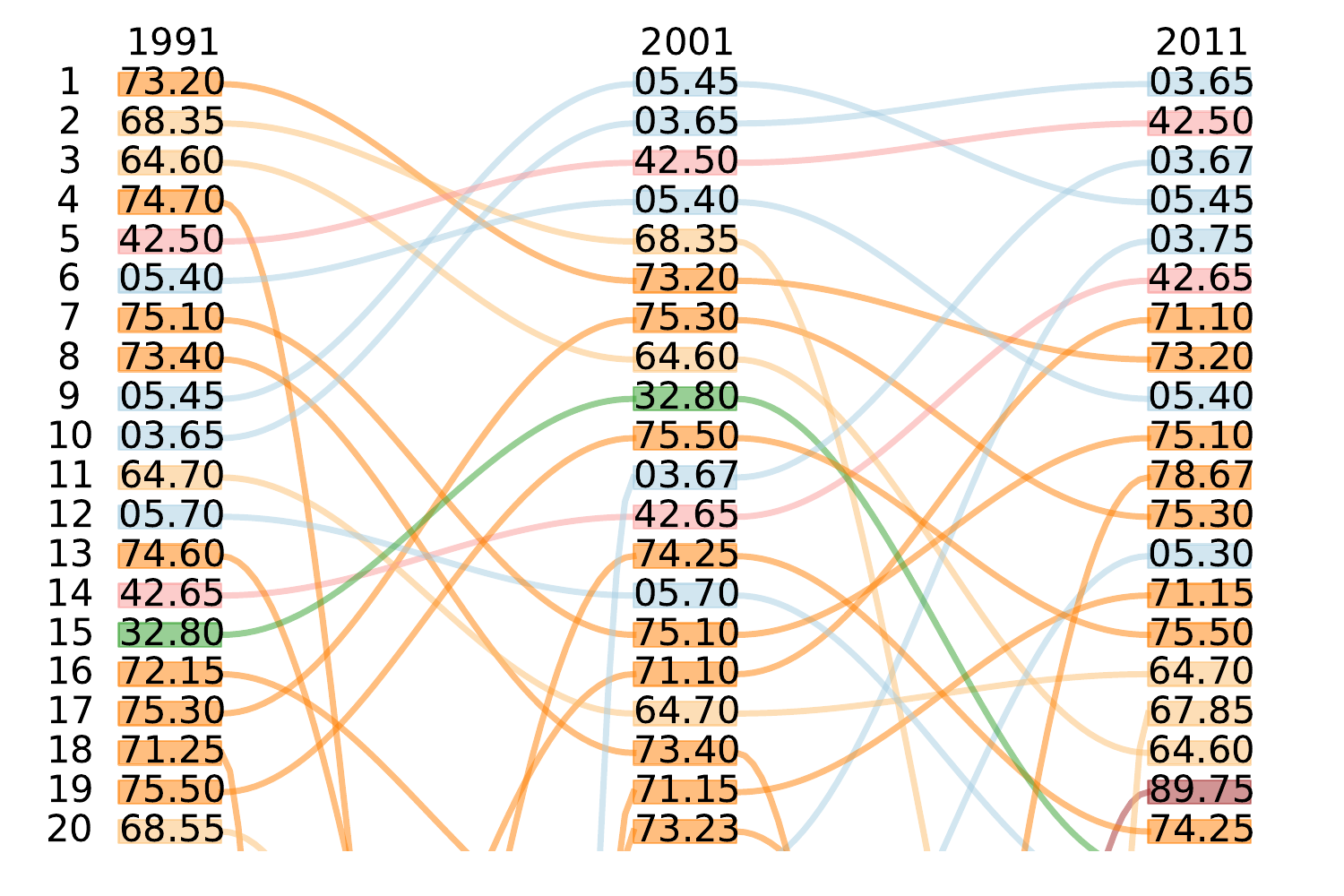}
\caption[Evolution of the top $20$ most influential level-$3$ subfields]{Evolutions of top $20$ most influential level-$2$ subfields for the years 1991, 2001 and 2011. At this level of details, we can see even more clearly that overall influences of fields in 60 and 70 are stable or slightly decreasing compared to the year of 1991 while fields in 03 and 05 are increasing. Full data can be downloaded at \revision{\url{subfield_list_level3.txt}}.}
\label{fig:evolution20-level3}
\end{figure}

\begin{figure}[htp]
\centering
\includegraphics[scale=0.25]{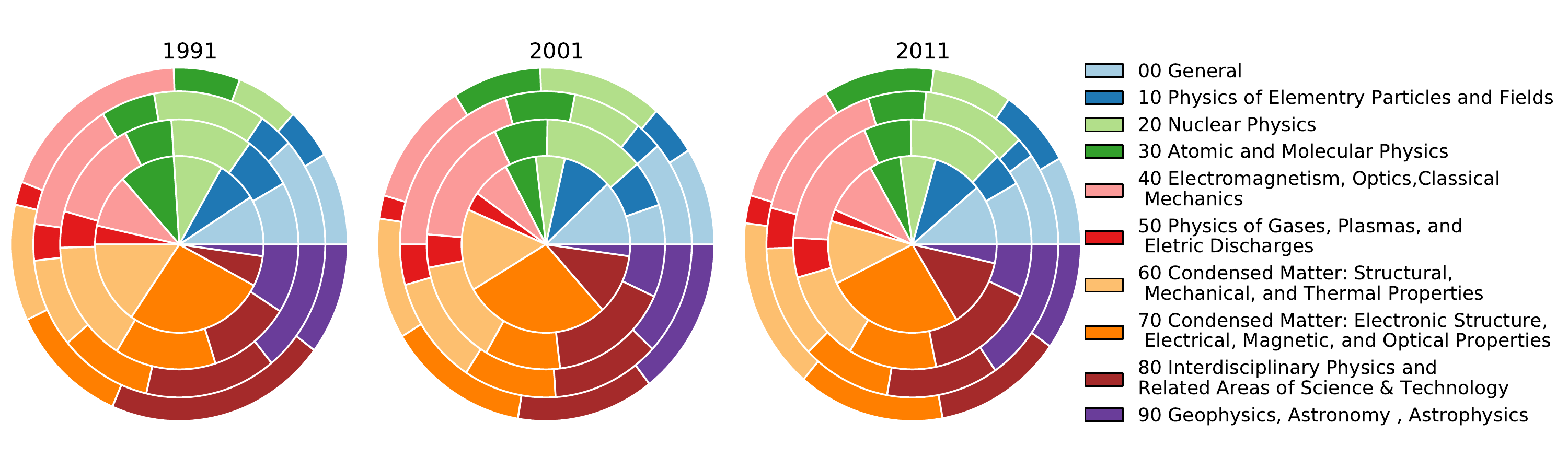}
\caption[Multi-level pie chart of major branches within each PACS region]{Multi-level pie charts 1991, 2001 and 2011 showing the composition of major branches of physics within each of the four $25\%$ regions. Colors represent the level-$1$ PACS codes, which are for the major branches of physics here. From the years, in the core region of the first quartile influential subfields, General Physics (00) and Interdisciplinary (80) have taken a larger part while Atomic and Molecular Physics has been shrinking.}
\label{fig:Pie}
\end{figure}

To look more closely into the finer structure, we analyzed the relative importances of the level-$3$ subfields and then plotted the level-$3$ subfields according to their level-$1$ classifications. Result of this analysis is presented using multi-layer pie charts in Fig.~\ref{fig:Pie}. Each level of the charts from inner to outer layers represents the ordered, from the most to the least influential ones, $25\%$ subfields. In each layer, we use different colors to represent the level-$1$ PACS codes of the subfields. Here level-$1$ PACS codes are regarded as the major branches of physics. In this way, we can see how each region is composed from major branches of physics and how this composition changed over time.

The pie chart for 1991 shows that the top quartile consisted mostly of Condensed Matter (PACS 60 and 70), General Physics (PACS 00) and Elementary Particles and Fields (PACS 01), with small contributions from Atomic and Molecular Physics (PACS 30), Electromagnetism, Optics, Classical Mechanics (PACS 40) and Interdisciplinary Physics (PACS 80). When the 2001 and 2011 pie charts are compared with the 1991 pie chart, we see General Physics and Interdisciplinary Physics become larger parts of the core region while Atomic and Molecular Physics shrinks. We can also look at the change in the distributions of a particular color between all four quartile sets. For example, Interdisciplinary Physics moves steadily toward the center, the more influential level whereas large portions of Geophysics, Astronomy and Astrophysics (PACS 90) migrate from the second quartile to third quartile most influential subfields.

\subsection{Figures for influences among subfields of physics}
In Fig. 6 in the main text, we choose two specific subfields at level-$2$, 98 and 03, and present subfields that are closely related to these two subfields. Here we report influences among all the level-$2$ subfields and also among all level-$1$ and level-$3$ subfields. We use a heatmap for this purpose: the size of each circle represents number of citations from the column subfield to the row subfield while the color in the circle corresponds to our IOI from the row to the column subfield. The value of the number of citations has been renormalized with respect to the row subfield.

First, we note that it is not always the case that the order of degree of influences is the same as the order of citation counts. Second, we found that a large numbers of IOIs are positive while a few of them are negative. We interpret the positive ones, which means that when field $i$ is removed from the whole discipline outcomes of field $j$ decreases, to be the relying-on relation and the negative ones,  which means that outcomes of field $j$ increases when field $i$ is removed, to be competitive or substitutive relation. A detailed examination of all those relations, which has not been done in this work, should be interesting. Third and finally, we also observed that in level-$2$ and level-$3$ heatmaps, overall there are relatively stronger correlations among the subfields within the same categories (the diagonal block elements) than that among the subfields across categories (the off-diagonal block elements). This means that the boundaries between different categories, represented by the hierarchical structure of the PACS codes, indeed meaningfully describe closely the interconnections among subfields.

\begin{figure}[htp]
\centering
\includegraphics[scale=0.23]{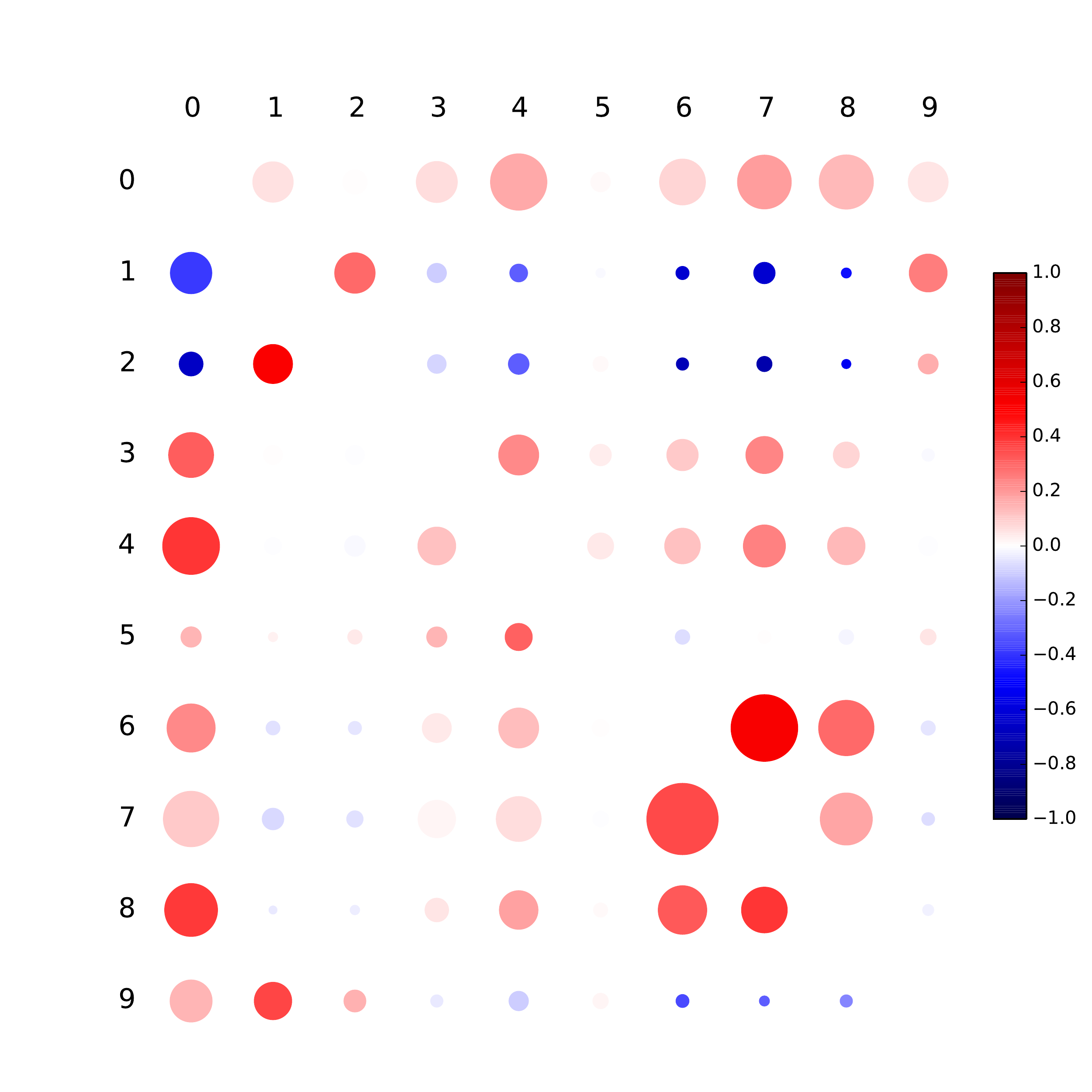}
\caption[Heatmap for influences among level-$1$ subfields]{Given a level-$1$ column subfield, its relations to other subfields in the row are color coded, as deeper color represents stronger influence. Size of the circles is proportional to the number of citations received by the row field from the column field. Full data can be downloaded at \revision{\url{heatmap_level1.txt}}.}
\label{fig:heat1}
\end{figure}

\begin{figure}[htp]
\centering
\includegraphics[scale=0.23]{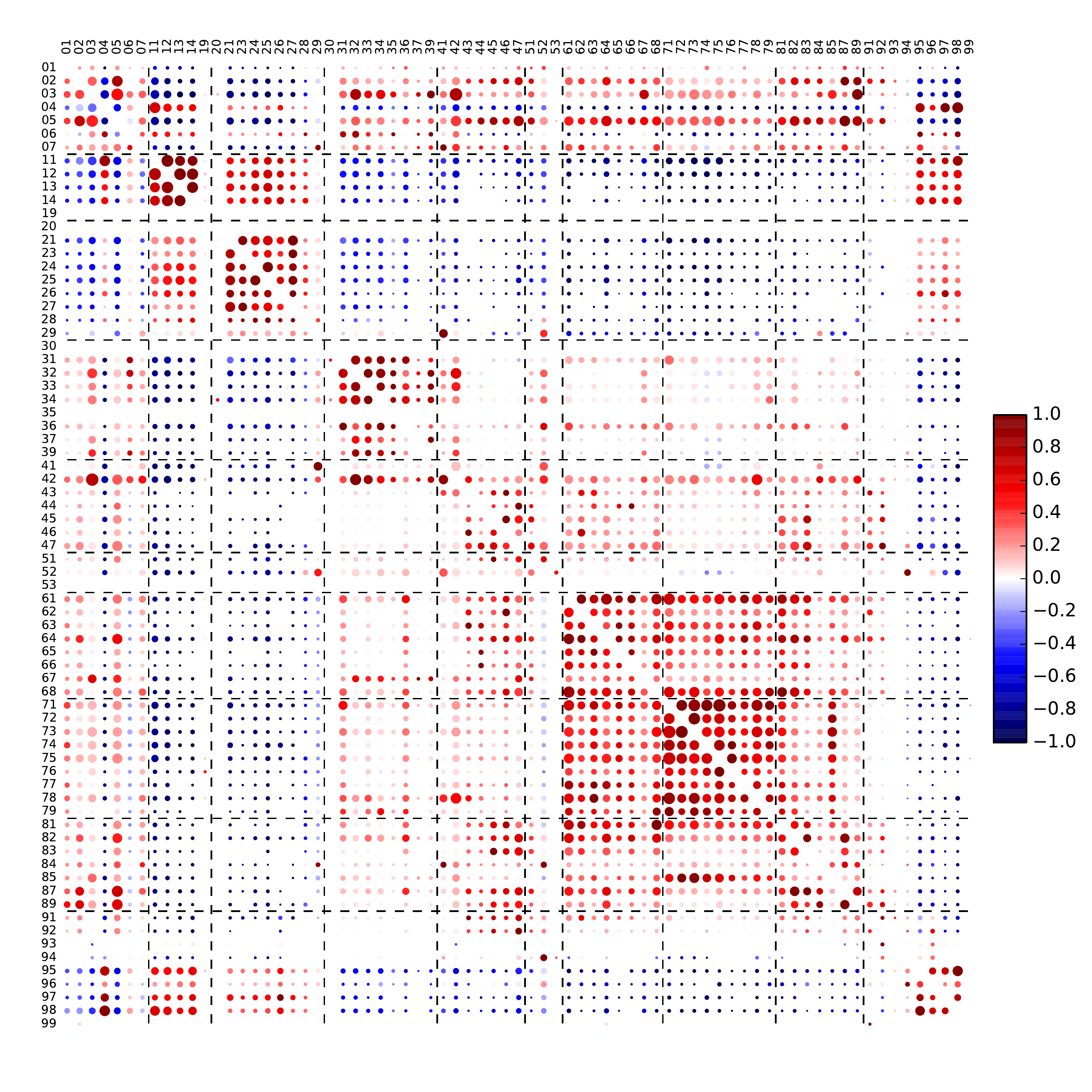}
\caption[Heatmap for influences among level-$1$ subfields]{Given a level-$2$ column subfield, its relations to other subfields in the row are color coded, as deeper color represents stronger influence. Size of the circles is proportional to the number of citations received by the row field from the column field. Full data can be downloaded at \revision{\url{heatmap_level2.txt}}.}
\label{fig:heat2}
\end{figure}


The level-$3$ heat map has $940\times940$ entries, so it is too big to show the figure in great detail. Therefore, for this map, we also provide a png file for downloading \url{relation_heatmap_level3.png} and a data file in excel format, which is accessible via \url{heatmap_level3.txt} (also \url{heatmap_level1.txt} and \url{heatmap_level2.txt} for the level-$1$ and level-$2$ heatmaps respectively). 
